\documentclass[a4paper,fleqn,usenatbib]{mnras}

\usepackage[pdftex]{graphicx}
\usepackage{eurosym}
\usepackage{amsmath}
\usepackage{multirow}
\usepackage{amsfonts}
\usepackage{adjustbox}
\usepackage{placeins}

\usepackage[T1]{fontenc}
\usepackage{ae,aecompl}

\usepackage{graphicx}	
\usepackage{amsmath}	
\usepackage{amssymb}	

\def\CN2{\mbox{$C_N^2 \ $}}
\def\CT2{\mbox{$C_T^2 \ $}}
\def\tauO{\mbox{$\tau_{0} \ $}}
\def\thetaO{\mbox{$\theta_{0} \ $}}

\title[OT forecast: ready for an operational application]{Optical turbulence forecast: ready for an operational application}
\author[E. Masciadri et al.]{E. Masciadri$^{1}$\thanks{E-mail:
     masciadri@arcetri.astro.it}, F. Lascaux$^{1}$, A. Turchi$^{1}$, L. Fini$^{1}$\\ 
$^{1}$INAF Osservatorio Astrofisico di Arcetri, Largo Enrico Fermi 5, I-501 25 Florence, Italy}

\date{Accepted 2016 November. Received 2016 November; in original form 2016 September}

\pubyear{2016}

\begin{document}
\label{firstpage}
\pagerange{\pageref{firstpage}--\pageref{lastpage}}
\maketitle

\begin{abstract}
One of the main goals of the feasibility study MOSE (MOdelling ESO Sites) is to evaluate the performances of a method conceived to forecast the optical turbulence above the ESO sites of the Very Large Telescope and the European-Extremely Large Telescope in Chile. The method implied the use of a dedicated code conceived for the optical turbulence (OT) called Astro-Meso-Nh. In this paper we present results we obtained at conclusion of this project concerning the performances of this method in forecasting the most relevant parameters related to the optical turbulence ($\CN2$, seeing $\varepsilon$, isoplanatic angle $\theta_{0}$ and wavefront coherence time $\tau_{0}$). Numerical predictions related to a very rich statistical sample of nights uniformly distributed along a solar year and belonging to different years have been compared to observations and different statistical operators have been analyzed such as the classical bias, RMSE, $\sigma$ and more sophisticated statistical operators derived by the contingency tables that are able to quantify the score of success of a predictive method such as the percentage of correct detection (PC) and the probability to detect a parameter within a specific range of values (POD). The main conclusions of the study tell us that the Astro-Meso-Nh model provides performances that are already very good to definitely guarantee a not negligible positive impact on the Service Mode of top-class telescopes and ELTs. A demonstrator for an automatic and operational version of the Astro-Meso-Nh model will be soon implemented on the sites of VLT and E-ELT.
\end{abstract}



%
%
\section{Introduction}

The MOSE (MOdelling ESO Sites) project is a feasibility study whose principal goal is to prove the possibility to forecast all the most relevant classical atmospheric parameters for astronomical applications (wind speed intensity and direction, temperature and relative humidity) and the optical turbulence (OT) that means $\CN2$ profiles with the integrated astroclimatic parameters derived from the $\CN2$ (i.e. seeing $\varepsilon$, isoplanatic angle $\theta_{0}$ and wavefront coherence time $\tau_{0}$) above the two European Southern Observatory (ESO) sites of Cerro Paranal, the site of the Very Large Telescope (VLT) and Cerro Armazones, the site of the European Extremely Large Telescope (E-ELT). The ultimate goal of the project is to investigate the possibility to implement an automatic and operational system for the forecast of these parameters at the VLT and at the E-ELT. In this paper we will treat the optical turbulence that, as will see later on, is the most difficult but also the most challenging parameter to be forecasted among those we treated. In previous papers related to the MOSE project we treated the abilities of the model in reconstructing wind speed and direction, temperature and relative humidity all along the atmosphere ($\sim$ 20 km i.e. vertical stratification of the atmospheric parameters) \citep{masciadri2013} and in reconstructing the same atmospheric parameters close to the surface \citep{lascaux2013,lascaux2015}. In all cases results indicated excellent model performances. In this paper we will focus our attention on results we obtained on the analysis of the model performances in reconstructing the optical turbulence i.e. the astroclimatic parameters ($\CN2$ profiles, the seeing $\varepsilon$, the isoplanatic angle $\theta_{0}$ and the wavefront coherence time $\tau_{0}$) that are commonly used to optimized the observations supported by the adaptive optics (AO). This paper completes therefore the whole feasibility study carried out for ESO. 

The forecast of the optical turbulence is crucial for the success of the new generation telescopes. We refer the reader to the Introduction of \cite{masciadri2013} for a detailed description of the scientific challenges related to the forecast of the optical turbulence. We remind here the key elements. The traditional scheduling of scientific program using as a criterion only the quality of the scientific program has, indeed, important drawbacks and limitations. It has been widely accepted by the astronomical community that we had to take into account simultaneously the quality of the scientific program but also the status of the optical turbulence to optimize the use of the telescope otherwise we risk that the most challenging scientific programs are not carried out and the most important potentials of telescopes and instrumentation are not exploited as they could. The forecast of the OT is therefore extremely important to schedule the scientific programs, to select the typology of instruments to be used at a specific time of the night and to optimize the adaptive optics (AO) systems performances. The AO techniques can be very powerful, at present, in correcting the perturbations induced by the optical turbulence on the wavefronts but the AO performances are strongly dependent on the status of the optical turbulence and, under particular conditions, they can hardly run or they can even not run at all. More in general the forecast of the optical turbulence is crucial for the Service Mode i.e. the observation mode of all the top-class facilities of present time and it will be the observing mode of all new generation facilities. It is the observing strategy that will maximize the possibility to achieve outstanding scientific goals with the ELTs. The Service Mode implies the knowledge in advance of the status of the atmosphere (atmospheric parameters and optical turbulence) and the rank of the scientific programs. The {\it forecast} of these parameters plays therefore a crucial role in the context of the high-angular resolution ground-based astronomy. Besides, we have not to forget that the cost of a night of observations is of the order of a hundred thousand US Dollars and it is therefore immediate to understand that the forecasts plays a crucial role not only in scientific but also in economical terms. The forecast we are dealing about aims to provide information in advance on a time scale $\Delta$T that is not inferior to 20 minutes. This is the typical time required to a beam to be shifted from an instrument to another in a configuration of permanents instruments placed in different focal stations. This is the configuration planned for new generation telescopes.

Our approach implies the use of atmospheric non-hydrostatic mesoscale models, more precisely a model called Meso-Nh \citep{lafore1998} for the atmospheric parameters joint with a dedicated code developed for the optical turbulence \citep{masciadri1999}. For simplicity we call this model Astro-Meso-Nh model. We refer the reader to the Introduction of \cite{masciadri2013} to know why mesoscale models are necessary instead of other typologies of models (General Circulation Models (GCM), Direct Numerical Simulations (DNS), Large Eddy Simulations (LES)). Mesoscale model are applied on limited areas of the Earth. There are different typologies of mesoscale models depending on the typical extension of the limited area and the horizontal resolution used. We used here limited areas having a size between 800 and 10 kilometers square and a subkilometric horizontal resolution in the innermost domain in the neighboring of the site of interest. The optical turbulence (OT) is completely parameterized in the mesoscale models. These characteristics guarantee to the Astro-Meso-Nh model to reconstruct the OT maintaining the link with the spatio-temporal evolution of the atmospheric flow external to the limited areas i.e. to realize a real 'forecast' of the OT. 

The Astro-Meso-Nh model has been applied in the last decades to many among the most important astronomical sites such as Cerro Paranal in Chile \citep{masciadri1999b}, San Pedro M\'artir in Mexico (\citealt{masciadri2002,masciadri2004,masciadri2006}), Roque de los Muchachos in Canaries Islands \citep{masciadri2001b}, Mt. Graham in Arizona \citep{hagelin2011}, Dome C in Antarctica (\citealt{lascaux2009,lascaux2010,lascaux2011}). For completeness we remind that other studies concerning the OT forecast on the whole atmosphere have been carried out using other mesoscale or general Circulation Models and similar (or different) approaches in the astronomical context (\citealt{cherubini2011,ye2011,giordano2013,liu2015}). 

We highlight three important considerations: {\bf (1)} the paper contains necessarily only a selection of the most relevant results obtained for the OT related to an extended study lasted a few years (and completed recently) that provided a clear indication of the good efficiency of the Astro-Meso-Nh model for an application to the Service Mode. These convincing results have induced ESO to propose us to implement a demonstrator for an automatic operational version of the Astro-Meso-Nh model applied to the sites of the VLT and the E-ELT. This project will start in the next months. Even if the research on the 'OT forecast' is in a continuum evolution (as well as that of the 'weather forecast') and there is always space for improvements, in this feasibility study we achieved important steps ahead in terms of estimation of the model performances. Due to the fact that we are entering in the new phase of the operational demonstration and more and more Observatories are interested on such a kind of application it is important to provide the state of the art of the performances of our system. {\bf (2)} Thanks to the development of our most recent algorithm of the $\CN2$ we could prove to be able to achieve a vertical resolution of the $\CN2$ all along the whole atmosphere up to roughly 150 m. This is a crucial achievement that opens interesting new perspectives for the most sophisticated adaptive optics systems i.e. the wide field adaptive optics (WFAO) such as the Ground Layer Adaptive Optics (GLAO) \citep{rigaut2002}, Multi Conjugated Adaptive Optics (MCAO) (\citealt{beckers1988,johnston1994}), Laser Tomography Adaptive Optics (LTAO) \citep{foy1985} and Multi Objects Adaptive Optics (MOAO) \citep{assemat2003}.  {\bf (3)} As expected, the model performances in forecasting the optical turbulence are not as good as in forecasting the atmospheric parameters (at least so far). This is due to the fact that the spatio-temporal scales on which the OT fluctuates are much smaller than the grid-size and also to the fact that the turbulence is a stochastic quantity. This means that it is more difficult to describe numerically the optical turbulence. However, in spite of these intrinsic difficulties, we will see that results we achieved are very impressive and, even more important, are objectively already of great support for the Service Mode. 

The plan of the paper is the following: in Section \ref{obs} we will describe the observations that we used as a reference to calibrate and to validate the model. In Section \ref{astro-meso-nh} we will described the model configuration used for this study. In Section \ref{method} we will describe the strategy used to calibrate and validate the model and in Section \ref{results} we will show the results obtained. In Section \ref{cn2_hvr} we show the model performances in reconstructing $\CN2$ profiles with very high vertical resolution. In Section \ref{concl} we will present the conclusions and  perspectives of this study.
%
%
\section{Observations}
\label{obs}

Measurements provided by different instruments have been used to carry out this study. Considering the scarcity of OT measurements related to Cerro Armazones, in agreement with ESO the study on the optical turbulence has been performed only above Cerro Paranal. An important preliminary analysis having the goal to assure and test the reliability of the measurements has been performed with part of the instruments of the PAR2007 site testing campaign \citep{dali2010} useful in our context: (1) a generalized-SCIDAR (more precisely the CUTE-SCIDAR III\footnote{Hereafter we will call the generalized-SCIDAR 'CUTE-SCIDAR III'  simply GS.}) developed by the Istituto de Astrofisica de Canarias (IAC) team \citep{vazquez2008} and corrected by \citet{masciadri2012} to eliminate the error induced by the normalization of the autocorrelation of the scintillation maps by the autocorrelation of the mean pupil (problem identified by \cite{johnston2002} and \cite{avila2009}), (2) a Multi-Aperture Scintillation Sensor (MASS) developed by the Kornilov \& Tokovinin team \citep{kornilov2003} and (3) a Differential Image Motion Monitor (DIMM) an instrument that since 1988 is running at Cerro Paranal to monitor the seeing i.e. the integration of the optical turbulence developed all along the whole atmosphere \citep{sarazin1990}. We had simultaneous GS and DIMM measurements related to 20 nights. MASS measurements were simultaneous to GS on a sub-sample of 14 nights. We note that the three instruments are located at basically the same height (GS at 5~m above ground level (a.g.l), DIMM and MASS at 6~m a.g.l.). Moreover the VLT is basically a plateau and this guarantees a fair comparison between measurements. 
The re-calibration of the GS measurements was fundamental to assure us a reliable reference. As we will see later on, we decided to use the GS as a reference for the calibration of the Astro-Meso-Nh model. Besides, a detailed analysis of comparisons of measurements from the GS, MASS and DIMM \citep{masciadri2014} permitted us to conclude that the MASS could not be taken as a reference because of three main problems: {\bf (1)} it underestimates the integrated turbulence (J or seeing) in the free atmosphere with respect to the GS with a relative error of -32$\%$ in terms of the seeing (-48$\%$ in terms of J); {\bf (2)} we found important discrepancies between MASS and GS in all the individual layers (reaching relative errors as high as -65\% in terms of seeing and -82\% in terms of J) with exception of the layers located at 2 and 16 km (layers 3 and 6) in which the relative error remains limited to +18\% in terms of seeing (+20\% in terms of J). A previous study on a similar topic \citep{tokovinin2005} (even if it was applied on a poorer statistical sample) found relative errors on individual layers as large as those we found in \cite{masciadri2014}; {\bf (3)} the particular weighting functions (WFs) of the MASS, having a triangle shape, do not permit to identify precisely the height of the boundary layer and in general the height separating a layer from the contiguous one. This represents an important limitation for the calibration of the Astro-Meso-Nh model. 

Thanks to DIMM measurements (a third independent instrument during the PAR2007 campaign together with GS and MASS) we could prove that the problem causing the discrepancies between GS and MASS came from the MASS. We cite here just the elements useful to justify why MASS measurements could not be used to calibrate the Astro-Meso-Nh model\footnote{We refer the reader to \citet{masciadri2014} for further details/discussion on the GS/MASS comparison. This is not part of the content of this paper.}. An accurate estimate of the $\CN2$ is indeed important for the model calibration and, because of the reasons we have just discussed, the MASS could not assure that.

Besides, we remind that \cite{masciadri2014} proved that the isoplanatic angle $\theta_{0}$ coming from the MASS is reliable (mainly thanks to the good reliability of the OT estimate in layer 6 located at 16 km above the ground) and the $\tau_{0}$ measurements are reliable too (just a few warnings with respect to this parameter - see the cited paper). This information is useful for the analysis done in this paper.

With these elements in mind we concluded we could use MASS measurements of $\theta_{0}$ and $\tau_{0}$ to validate the model with respect to these parameters (see Section \ref{cal_val}). MASS is, indeed, an instrument currently running above Cerro Paranal at VLT Observatory. After the model calibration, we could therefore validate the model with a more extended sample of nights not belonging to the PAR2007 campaign.

\begin{table}
\caption{Astro-Meso-NH model grid-nesting configuration for the OT simulation.
In the second column the number of horizontal grid-points, in the third column the domain extension and
in the fourth column the horizontal resolution $\Delta$X.}
\begin{center}
\begin{adjustbox}{max width=\columnwidth}
\begin{tabular}{|c|c|c|c|}
\hline
Domain & Grid   & Domain size & $\Delta$X  \\
       & Points & (km)        & (km)  \\
\hline
Domain 1 &  80$\times$80  & 800$\times$800 & $\Delta$X = 10\\
Domain 2 &  64$\times$64  & 160$\times$160 & $\Delta$X = 2.5\\
Domain 3 & 150$\times$100 &  75$\times$50  & $\Delta$X = 0.5\\
\hline
\end{tabular}
\end{adjustbox}
\label{tab:gn_config1}
\end{center}
\end{table}

\begin{figure}
\centering
\begin{adjustbox}{max width=\columnwidth}
\includegraphics[width=0.9\columnwidth]{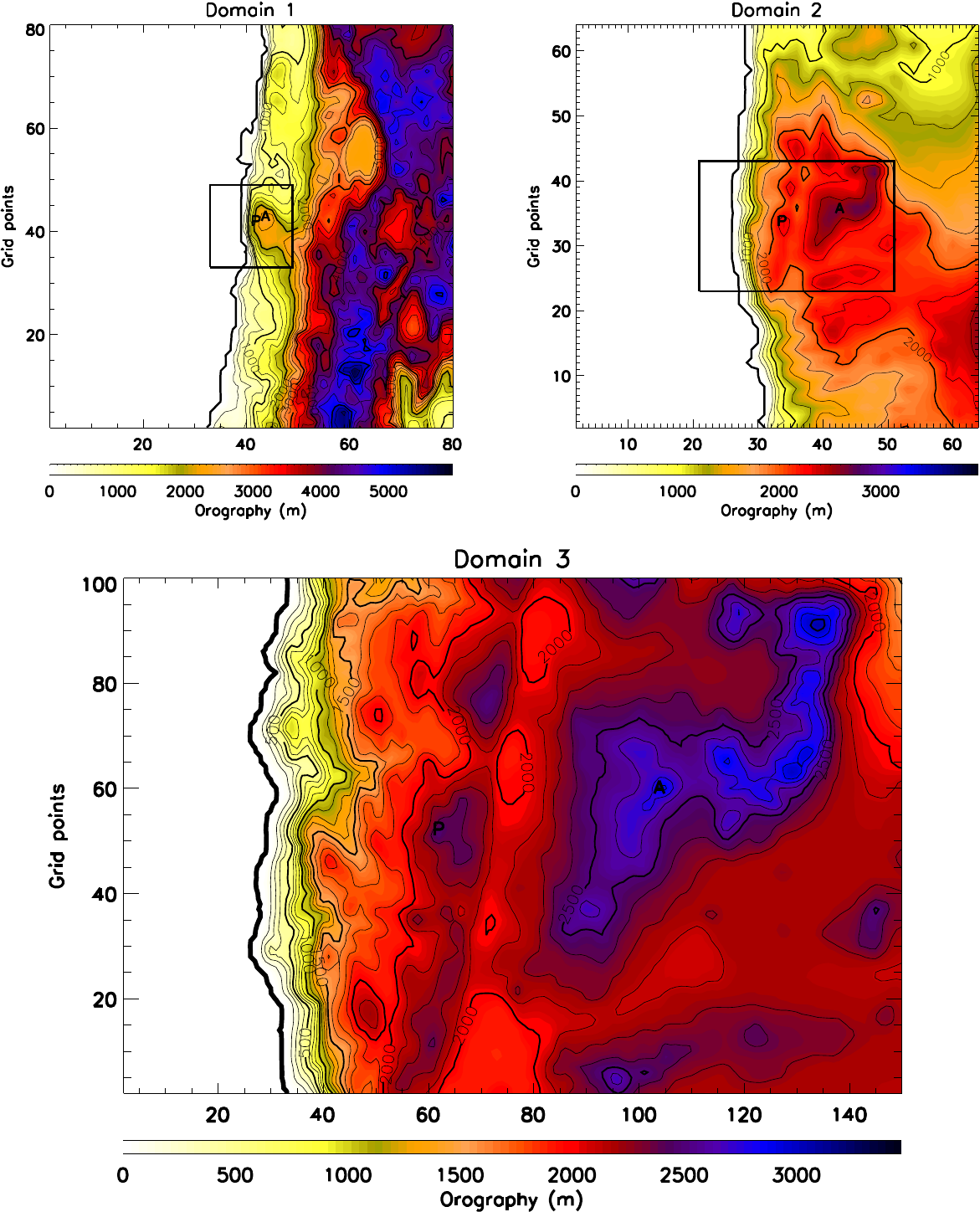} 
\end{adjustbox}
\caption{Orography (altitude in m) of the region of interest as seen by the Meso-NH model for all the embedded domains of the model grid-nested configuration.
{\bf (a)} Domain 1 (digital elevation model i.e. orography data from GTOPO),
{\bf (b)} Domain 2 (digital elevation model from GTOPO),
{\bf (c)} Domain 3 (digital elevation model from ISTAR),
'P' stands for Cerro Paranal, 'A' stands for Cerro Armazones. The black square in Domain 1 represents the surface of Domain 2. The black rectangle in Domain 2 represents the surface of Domain 3.
See Table~\ref{tab:gn_config1} for the specifications of the domains (number of grid-points, domain extension, horizontal resolution).}
\label{orog}
\end{figure}

\clearpage

%
%
\section{Model configuration}
\label{astro-meso-nh}

As previously said we used the Meso-Nh model\footnote{\url{http://mesonh.aero.obs-mip.fr/mesonh52} - we used the Masdev4.8 version of the code.} interfaced with the Astro-Meso-Nh code. Both models/codes can simulate the temporal evolution of three, two or mono-dimensional parameters over a 
selected limited area of the globe. In the Astro-Meso-Nh code, the $\CN2$ is a 3D parameter, all the integrated astroclimatic parameters ($\varepsilon$, $\theta_{0}$ and $\tau_{0}$) are 2D parameters. The two codes are run together. For this reason we provide the configurations of both models. For what concerns Meso-Nh the system of hydrodynamic equations is based upon an anelastic formulation that permits an effective filtering of acoustic waves. The model uses the \cite{gal1975} coordinates system on the vertical and the C-grid in the formulation of \cite{arakawa1976} for the spatial digitalization. The model version used for this study employs an explicit three-time-level leap-frog temporal scheme with a time filter \citep{asselin1972}. The model employs a one-dimensional 1.5 turbulence closure scheme \citep{cuxart2000} and we used a one-dimensional mixing length proposed by \cite{bougeault1989}. The surface exchanges are computed using the interaction soil biosphere atmosphere (ISBA) module \cite{noilhan1989}. 
The OT and derived integrated parameters are simulated with the Astro-Meso-Nh code developed by \cite{masciadri1999} and since there in continuous development by our group. The geographic coordinates of Cerro Paranal are (24$^{\circ}$37'33.117" S, 70$^{\circ}$24'11.642 W). We used the grid-nesting techniques \citep{stein2000} consisting in using different embedded domains of the digital elevation models (DEM i.e. orography) extended on smaller and smaller surfaces, with increasing horizontal resolution but with the same vertical grid. We employed the two-way grid-nesting that guarantees a better thermodynamic balance of the atmospheric flow because it takes into account the feedback between each couple of father and son domain. Gravity waves propagation is also better reconstructed than the one-way option. Simulations of the OT consisted on three embedded domains where the horizontal resolution of the innermost domain was $\Delta$X = 500~m (Table \ref{tab:gn_config1}). Fig.~\ref{orog} shows the three domains as seen by model. The model is initialized with analyses provided by the General Circulation Model (GCM) HRES of the European Center for Medium Weather Forecast (ECMWF) having an intrinsic horizontal resolution of 16 km\footnote{Starting from March 2016 ECMWF products have an horizontal resolution of roughly 9~km.}. All simulations we performed with Astro-Meso-Nh start at 18:00 UT of the day before (J-1) and they last in total 15 hours. We reject the first 6 hours from our analysis because we are interested on data after the sunset and we therefore take into account Astro-Meso-Nh outputs starting from 00:00 UT of the day J. During the 15 hours the model is forced each six hours (synoptic hours) with the analyses provided by the GCM related to the correspondent hours. In the grid point of the summit of Cerro Paranal we print out the $\CN2$ profiles with a temporal sampling of 2 minutes from the innermost domain\footnote{The $\CN2$ is calculated obviously by the model each time step.}. We selected 2 minutes because this corresponds more or less to the sampling of typical vertical profilers. There are no major constraints in shortening it but, at present, neither a particular interest in doing that. The vertical grid is the same for all the domains. We have a 62 vertical levels with a first grid point of 5~m, a logarithmic stretching of 20 per cent up to 3.5~km above the ground and almost constant vertical grid size of $\sim$600~m up to 23.8~km. The heigh of the first grid point is determined by the height above the ground at which the Generalized SCIDAR, that we used to compare measurements and simulation, is located (5~m). It is important to highlight that this model configuration guarantees the implementation of the model in an operational set-up. 
\begin{figure*}
\begin{center}
\begin{tabular}{c}
\includegraphics[width=0.95\textwidth]{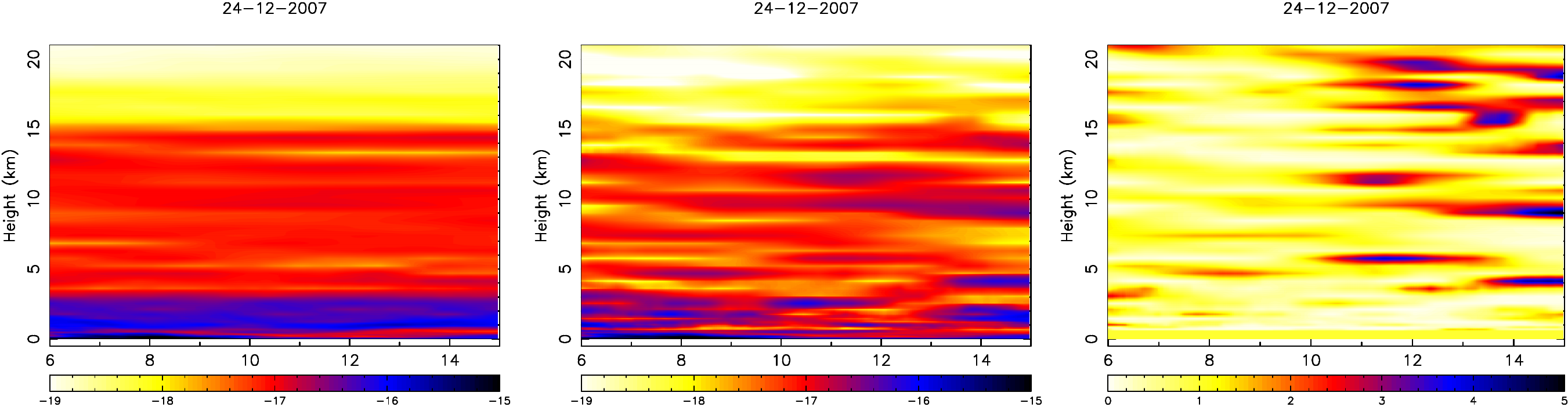}
\end{tabular}
\end{center}
\caption{\label{fig:corr_coef} Temporal evolution of the $\CN2$(z) during the night of 24/12/2007 in its standard form (Eq.\ref{eq:cn2_2}) (top left) and with the new algorithm (Eq.\ref{eq:cn2_stand1}, Eq.\ref{eq:cn2_stand2} and Eq.\ref{eq:cn2_mod_1})  (top center). On top right is shown the temporal evolution of the factor $A_{2}$. The x-axis report the time in hours from 00:00 UT (six hours after the initialization at 18:00 UT i.e. more or less the beginning of the night). The temporal interval shown in the three figures corresponds to the local night. For all figures a moving average of 1 hour has been applied.}
\end{figure*}

%
%
\section{Method of analysis} 
\label{method}

\subsection{Astro-Meso-Nh model calibration and validation}
\label{cal_val}

We considered in total three completely independent samples of nights (20, 36 and 53 nights). The 20 nights of the PAR2007 campaign (all concentrated in the months of November and December 2007) have been used for the model calibration because we had GS and DIMM simultaneous measurements. Only measurements from the GS have been used in reality to calibrate the model but DIMM measurements have been very useful to assure us the reliability of the GS measurements (see Section \ref{obs}). To calibrate the Astro-Meso-Nh model we need a vertical stratification of the OT i.e. $\CN2$ profiles. For the model validation we could no use GS measurements since no measurements on long sample size are available for Cerro Paranal. We therefore considered for the model validation simulations done in the period [06/2010 - 05/2011; a solar year] for which DIMM and MASS measurements of the seeing were available. MASS and DIMM are routinely used by ESO to monitor atmosphere at Cerro Paranal. For the validation sample, for each month we selected the nights corresponding to the 1st, 5th, 15th and 25th of the month. In case the constraint was not respected we looked for the nearest night to the sequence reported above respecting this criterion. A sample of 53 nights has been selected. A second set of 36 nights has been selected in 2007 (but completely independent from the calibration sample of 20 nights of the PAR2007 campaign used for the model calibration) to investigate a specific effect observed in results obtained in our analysis of the 53 nights\footnote{The period in which we selected the 53 nights of the the final validation was more recent (2010-2011) than that related to the calibration sample (2007). 
We decided therefore to perform a validation of the model also on a second sample of 36 nights (independent from the calibration sample) 
but belonging to the same year (2007). This exercise to be sure to exclude effects due to the initialization data. }. 

The technique of the Astro-Meso-Nh model calibration has been proposed by \cite{masciadri2001} and validated later on by \cite{masciadri2004}. The calibration aims at fixing the value of the minimum turbulent kinetic energy (E$_{min}$) to be given at each fixed vertical slab of the model (or each vertical model level)\footnote{We refer the reader to \cite{hagelin2011} for a discussion on the difference between the two approaches.}. 
The E$_{min}$ is the minimum turbulent kinetic energy required by the model to move from its state of equilibrium. 
Once the E$_{min}$ value is fixed (result of the calibration), all the simulations are run again with this E$_{min}$ (the same for all simulations). E$_{min}$ can be considered a sort of background climatologic noise. In regions in which the dynamic turbulence is well developed the model rapidly forgets the E$_{min}$ (more frequent in the low part of the atmosphere). However, in stable regions of the atmosphere, \cite{masciadri2001} proved that $\CN2$ $\propto$ E$_{min}$$^{2/3}$ therefore E$_{min}$ can be retrieved through the minimization of the difference of observed and simulated $\CN2$ profiles (see Eq.8 in \cite{masciadri2001}). In each $\CN2$ profile typically we can identify more than one vertical slab in which turbulence is in stable regimes (see \cite{masciadri2001} - Fig.1). Different values of E$_{min}$ means we are assuming that the atmosphere has more or less inertia in different regions of the atmosphere and the thermodynamic instabilities require more or less energy to trigger turbulence in some regions of the atmosphere. 
Supported by these arguments the model calibration proposed by \cite{masciadri2001} has been applied with some different variants, such as the number and the extension of the verticals slabs, in several studies (\citealt{masciadri2004,masciadri2006,cherubini2011,hagelin2011}).
In this study we minimized the number of the vertical slabs so to get the model less dependent on the size of the calibration sample.
We divided the 20~km in just three vertical slabs: [0-600~m], [600~m-14~km] and [14~km-20~km]. The first threshold ($\sim$600~m) discriminates regions of the atmosphere in which the model is more or less active and weakly/strongly dependent from the calibration procedure. The second threshold ($\sim$14~km) corresponds, in average, to the height at which there is a more rapid change of the structure of the $\CN2$ at low spatial frequencies. This is also due to the fact that the stratosphere is characterised by different dynamic of the atmosphere than the troposphere. 
After the calibration, inside these three vertical slabs, the value of the E$_{min}$ assumes three different constant values: E$_{min,1}$, E$_{min,2}$ and E$_{min,3}$. 

In conclusion, the strategy of our analysis is therefore summarised in the following sequence of steps: (1) we perform the model calibration using the GS measurements; (2) we fix the free parameters E$_{min,i}$ in the model; (3) we repeat the whole set of simulations on the calibration sample with the same E$_{min,i}$; (4) we perform a preliminary validation i.e. we quantify the model performances with respect to the sample of night used for the calibration; (5) we perform a model validation i.e. we quantify the model performances with respect to a totally independent sample of nights.
Step (4) revealed useful in some circumstances because, as we will see in Section \ref{results}, in coincidence of the calibration sample we could access, in some cases, to measurements taken simultaneously with different and independent instruments (GS, DIMM and/or MASS). This permitted us to better appreciate the model performances and model uncertainties. 
%
%
\subsection{C$_{N}^{2}$ algorithm}

The algorithm used for the $\CN2$ is described in Eq.\ref{eq:cn2_2}:

\begin{equation}
C_{N}^{2}(z)=0.58 \cdot \varphi _{3}(z)\left [ \frac{80\cdot 10^{-6}\cdot P(z)}{T(z)\cdot \theta(z) } \right ]^{2}
\cdot L(z)^{4/3}\cdot \left [ \frac{\partial \theta(z) }{\partial z} \right ]^{2}
\label{eq:cn2_2}
\end{equation}
\noindent
where P is the atmospheric pressure, T the temperature, $\theta$ the potential temperature, $\varphi _{3}$ is a dimensionless function depending on the thermal and dynamic stability of the atmosphere proportional to the inverse of the Prandtl number and introduced in Meso-Nh by \cite{redel_somm_1981} and L the mixing length (more precisely the BL89 mixing length defined as follow: at any level $z$ in the atmosphere a parcel of air of given turbulence kinetic energy $e(z)$ can move upwards ($l_{up}$) and downwards ($l_{down}$) before being stopped by buoyancy forces. These distances are defined by: 
\begin{equation}
\int_{z}^{z+l_{up}}\frac{g}{\theta _{v,ref}}\left ( \theta _{v}(z)-\theta _{v}(z^{'}) \right )dz^{'}=e(z)
\end{equation}
and
\begin{equation}
\int_{z-l_{down}}^{z}\frac{g}{\theta _{v,ref}}\left ( \theta _{v}(z^{'})-\theta _{v}(z) \right )dz^{'}=e(z)
\end{equation}
where L is defined as:
\begin{equation}
L=\left [ \frac{\left (l_{up}  \right )^{-2/3}+\left (l_{down}  \right )^{-2/3}}{2} \right ]^{-3/2}
\label{eq:lm}
\end{equation}
\noindent
where $\theta _{v}$ is the virtual potential temperature and $\theta _{v,ref}$ is $\theta _{v}$ of reference related to the anelastic system. 
It represents the hydrostatic conditions (where the density depends just on the height z). We will call hereafter this algorithm '$\CN2$ reference algorithm' that is very similar to that proposed by \cite{masciadri1999}. For completeness we refer to Annex \ref{annex_a} to discuss the few technical differences among the two, that are however irrelevant for the rest of the analysis.
To improve the peak-to-valley temporal evolution of some astroclimatic parameters (free-atmosphere seeing, isoplanatic angle), we corrected the $\CN2$ algorithm of Eq.\ref{eq:cn2_2} by a factor that takes into account the wind shear according to Eq.\ref{eq:cn2_stand1}, Eq.\ref{eq:cn2_stand2} and Eq.\ref{eq:cn2_mod_1}: 

\begin{flalign}
\label{eq:cn2_stand1}
&for\ z<700\ m,\ C_N^2(z)^{*}=C_N^2(z)\\
\label{eq:cn2_stand2}
&for\ z>700\ m,\ C_N^2(z)^{*}=A(z)\cdot C_N^2(z)\\
\nonumber
\end{flalign}
with
\begin{flalign}
A(z)=\left\{\begin{matrix}
A_1(z)=[(\frac{dV_x}{dz})^2+(\frac{dV_y}{dz})^2]^{\beta}/<[(\frac{dV_x}{dz})^2+(\frac{dV_y}{dz})^2]^{\beta}> \\ 
or\\ 
A_2(z)=[(\frac{dV}{dz})^2]^{\beta}/<[(\frac{dV}{dz})^2]^{\beta}> \\ 
\end{matrix}\right.
\label{eq:cn2_mod_1}
\end{flalign}
\noindent
with $\beta$ $\in$  [$\frac{1}{2}$,$1$,$\frac{3}{2}$]. $V_x(z)$ and $V_y(z)$ are the horizontal components of the wind speed.
$V$(z) is the module of the horizontal wind. A forthcoming technical paper focused on this new $\CN2$ algorithm will treat in detail the motivation for the modification of the $\CN2$ algorithm, scientific justifications for such a modified algorithm, impact on model performances and perspectives for further improvements. In this paper we intend to trace the state of the art of the model performances using its best configuration in application to the ground-based astronomy i.e. the most interesting aspect for astronomers. We limit us to say that the new algorithm improves the peak-to-valley spatio-temporal variability of the model during the night providing a better correlation with measurements. 

The threshold of 700~m was selected because in the low part of the atmosphere the problem of the spatio-temporal variability is absent. To test the sensibility of the new algorithm with respect to the selected threshold we used as a threshold 700 m and 600 m (close to the threshold in which model behaviour changed) and it has been verified that results differed by negligible quantities. This means that we are not very sensitive to the threshold within at least a hundred of meters. On a sub-sample of nights we tested $A_1(z)$ and $A_2(z)$ in the new $\CN2$ algorithm formulation and we found negligible differences on the correction factor. We chose to use $A_2(z)$. Moreover, the best fit was obtained with $\beta = \frac{3}{2}$. The whole analysis presented in this paper is based on this 'new $\CN2$ algorithm' in which all the $\CN2$ profiles have been corrected by $A_2(z)$ with $\beta = \frac{3}{2}$. To give an idea of effect of the factor A$_{2}$(z) Fig.\ref{fig:corr_coef} shows, in the case of one night, the temporal evolution of the $\CN2$ in its standard form (Eq.\ref{eq:cn2_2}) and after the correction (Eq.\ref{eq:cn2_stand1}, Eq.\ref{eq:cn2_stand2} and Eq.\ref{eq:cn2_mod_1}). In the same picture is also shown the temporal evolution of the factor A$_{2}(z)$.

%
%
\section{Results: Astro-Meso-Nh code performances} 
\label{results}

The statistical operators we used in our analysis are the bias, the root mean squared error (RMSE) and $\sigma$ that provide fundamental information on the systematic (bias) and statistical uncertainties (RMSE and $\sigma$). Bias and RMSE are defined as:\\
\begin{equation}
BIAS=\sum_{i=1}^{N}\frac{Y_{i}-X_{i}}{N}
\label{eq:bias}
\end{equation}
and
\begin{equation}
RMSE=\sqrt{\sum_{i=1}^{N}\frac{\left ( Y_{i}-X_{i} \right )^{2}}{N}}
\label{eq:rmse}
\end{equation}
where Y$_{i}$ and X$_{i}$ are the individual values reconstructed by the model and observed by the instrument, N is the number of times for which a couple (X$_{i}$,Y$_{i}$) is available with both X$_{i}$ and Y$_{i}$ different from zero. Starting from the bias and the RMSE it is possible to retrieve the bias-corrected RMSE i.e. the $\sigma$:\\

\begin{equation}
\sigma=\sqrt{\sum_{i=1}^{N}\frac{[(X_i-Y_i)-(\overline{X_i-Y_i})]^2}{N}}\\ =\sqrt{RMSE^2-BIAS^2}
\label{eq:sigma}
\end{equation}

that expresses the intrinsic uncertainty not affected by the bias. We did not use the correlation coefficient because we proved that this operator is not reliable and it can induce to misleading conclusions (see Annex \ref{annex_b}). Besides, in order to have a more practical estimate of the model score of success we calculated the 'contingency tables' from which we deduced more sophisticated statistical operators: the percent of correct detection (PC), the probability to detect a parameters within a specific range of values (POD$_{i}$) and the probability of extremely bad detection (EBD).  Contingency tables allow for the analysis of the relationship between two or more categorical variables. We refer the reader to \cite{lascaux2015} for an exhaustive discussion about the role of 'contingency table' of different sizes. We refer the reader to Annex \ref{annex_c} for a brief summary of key definitions of contingency tables, PC, POD and EBD to allow the reader to follow the analysis of this paper. 

It is also important to mention that we could prove (see \citealt{masciadri2013b} for this issue) that the model calibration for seeing and isoplanatic angle are {\it 'season dependent'}. With this we mean that the model should be calibrated with different data related to the two seasons: winter [April-September] and summer [October-March]. However, in our study the unique available GS measurements for the calibration refer to the site testing campaign (PAR2007) performed in November and December i.e. in summer. Therefore for the seeing and isoplanatic angle we report and discuss here only results we obtained for the validation sample in the summer time i.e. the reliable results. We assume that, using a sub-sample of SCIDAR measurements taken in winter time, the model will be able to be calibrated for this season\footnote{A site testing campaign covering a whole solar year will be done with a Stereo-SCIDAR in the near future at Cerro Paranal \citep{osborn2016}. We will be able therefore to perform a dedicated calibration for the winter time for the time in which we will implement the 'demonstrator' of the operational system for the optical turbulence forecast at the VLT}. The issue of the dependency of the model calibration from the season is almost not observed on the wavefront coherence time. We think that the reason is due to the fact that $\tau_{0}$ depends on two parameters: the $C_{N}^{2}$ and the wind speed and probably the good model reconstruction of the wind speed can mask and overcome somehow this effect.  

Even if the original temporal sampling of measurements and simulation is high (2 min is the temporal sampling of the $\CN2$ from which depends all the other parameters), all measurements and simulations of $\varepsilon$, $\theta_{0}$ and $\tau_{0}$ have been treated with a moving average of 1 hour plus a resampling 20 minutes before to calculate the scattering plots and the contingency tables. The moving average of 1 h has been chosen because astronomers are
more interested in the trend of the prediction. It has been observed that the moving average, smoothing out the high frequencies, is more efficient to identify the model and measurement trends than a simple resampling. The high-frequency variability on shorter time-scales is less relevant and useless in our context because the scheduling of scientific programs can not be tuned with such high frequency. Astronomers are interested in identifying whether the trend of a parameter is increasing, decreasing or is stationary, in order to be able to take a decision about changing a modality of observation or a scientific program. The selection of the interval of 20 min for the resampling of measurements is justified by the fact that this is more or less the effective time necessary to switch from one modality of observation to another. The values of 1 h and 20 min have been selected in agreement with ESO staff\footnote{We remind to the reader the importance to have for both, measurements and simulations, an original high temporal sampling. That is very different from having an output each hour that definitely does not permit to quantify a trend for a stochastic parameter such as the optical turbulence.}. 

We focused our attention on the analysis of the validation sample that is, obviously,  the most important. We reported in this section also some results obtained on the analysis of the calibration but just in some specific contexts when results provided interesting new insights that deserved a discussion. This is to avoid to scatter the attention from the main outputs of this study.

\begin{figure}
\begin{center}
\begin{tabular}{c}
\includegraphics[width=0.4\textwidth]{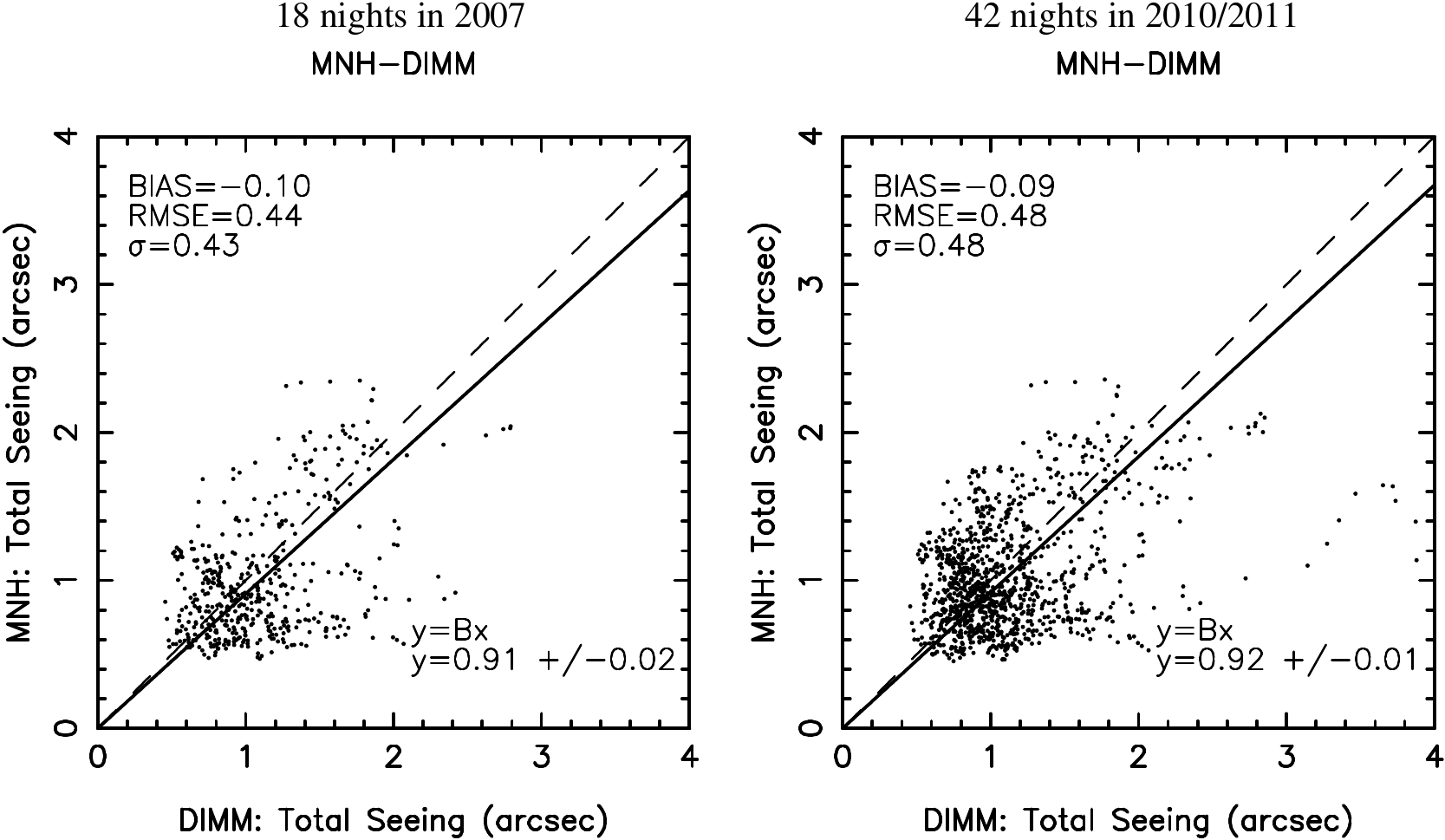}\\
\end{tabular}
\end{center}
\caption{\label{fig:clouds_seeing_mnh_dimm} Validation samples: scattered plot of the total seeing between Meso-NH outputs and DIMM measurements, for the
summer periods of 2007 (case B) on the left, and 2010/2011 (case A) on the right.}
\end{figure}
\begin{figure*}
\begin{center}
\begin{tabular}{c}
\includegraphics[width=0.70\textwidth]{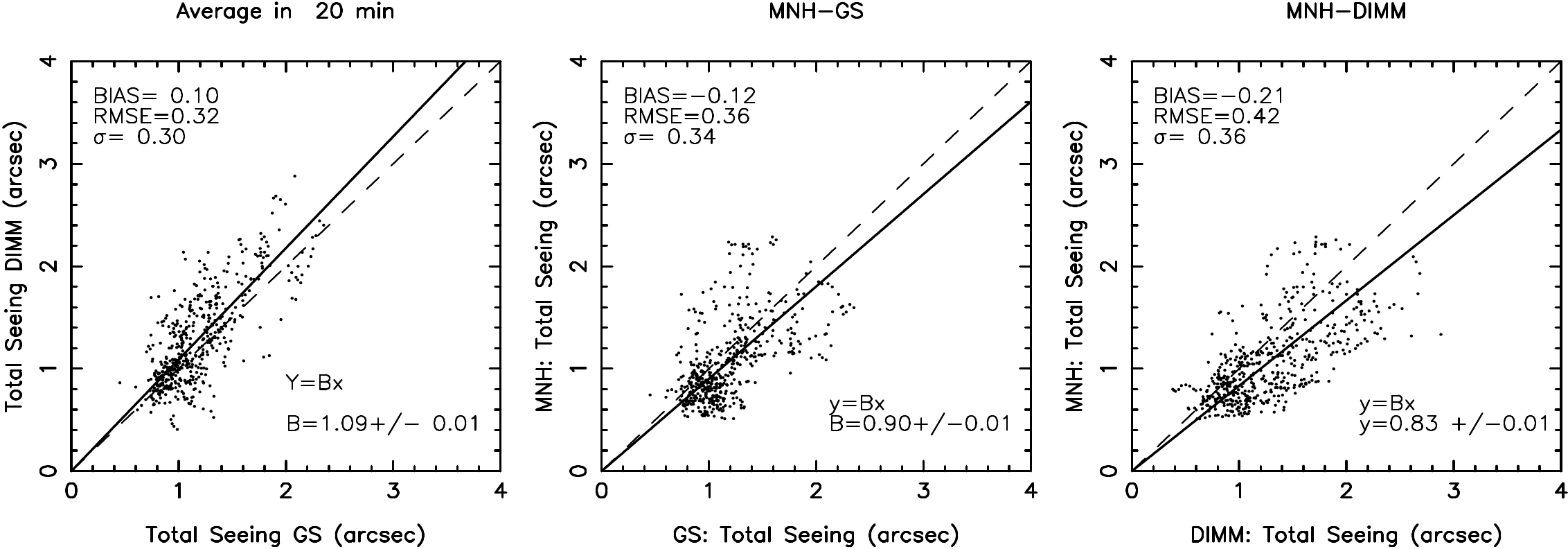}\\
\end{tabular}
\end{center}
\caption{\label{fig:clouds_seeing_mnh_dimm_gs_20} Scattered plot of the total seeing between Meso-NH outputs and DIMM and GS measurements (center and right) and between the GS and the DIMM (left) for the calibration sample of 20 nights.}
\end{figure*}

%
%
\subsection{Seeing - $\varepsilon_{0}$}
\label{seeing}
We analyze in this section 3$\times$3 contingency tables, with 2 different sets of thresholds.
In the first case (CASE 1), we use, as thresholds of the total seeing, the first and second tertiles of the cumulative distribution calculated on the sample of nights considered in the validation sample. 
In the second case (CASE 2), we use as thresholds of the total seeing, 1 arcsec and 1.4 arcsec.
A seeing weaker than 1 arcsec and larger than 1.4 arcsec should still provide a very useful information to help for the decision to be taken {\it in loco} and it can be considered a challenging achievement\footnote{The values of 1 and 1.4 arcsec have been decided after consultation with the ESO MOSE Board and colleagues from the adaptive optics group of the Arcetri Astrophysical Observatory.}. Only the DIMM is used as a reference when we treated the large validation sample for the seeing since it is the only instrument providing seeing values on the whole atmosphere. GS measurements are available only for the calibration sample (20 nights related to the PAR2007 campaign). We treated two samples: (A) the whole sample of 53 plus 36 nights i.e. 89 nights in 2010, 2011 and 2007 and (B) the sub-sample of 36 nights in 2007, the season for which the calibration of the model was done. Sample B is a sub-sample of sample A.

\subsubsection{Validation sample}
\label{val_see}

As explained before we focus our attention on the summer time (42 nights in the sample A and 18 in the sample B). Tables~\ref{tab:ct_seeing_val_2010_1} and \ref{tab:ct_seeing_val_2010_2} 
are the contingency tables for both cases (1 and 2), related to the 42 nights in summer time of sample A. Tables~\ref{tab:ct_seeing_val_2007_2} and \ref{tab:ct_seeing_val_2007_1} (in Annex \ref{annex_d}) are the contingency tables for both cases (1 and 2), related to the 18 nights in summer time of sample B. Figure~\ref{fig:clouds_seeing_mnh_dimm} displays the scattered plot of the total seeing between Meso-NH outputs and DIMM measurements, for the 18 nights of summer time of sample B (left), and that of the 42 nights of summer time of sample A (right). We can observe an almost negligible bias and a $\sigma$ of 0.43 arcsec and 0.48 arcsec. Looking at the contingency table just cited it is possible to see that the values of POD$_{i}$ are in general smaller than those we obtained for the atmospheric parameters (see \citealt{lascaux2015}) however the external POD$_{i}$ (POD$_{1}$ and POD$_{3}$) are well above the value of 33\% correspondent to the random case. From one side this tells us that, as expected, the prediction of astroclimatic parameters is more difficult. However, from an astronomical point of view it is much more critical to know in advance the probability to detect the smallest seeing values i.e. the most interesting and critical range is POD$_{1}$. POD$_{1}$ is the probability that the model predicts seeing weaker than the first threshold (case 1) or the probability that the model detect a seeing smaller than 1 arcsec (case 2).  In Tables~\ref{tab:ct_seeing_val_2010_1}, \ref{tab:ct_seeing_val_2010_2}, \ref{tab:ct_seeing_val_2007_2} and \ref{tab:ct_seeing_val_2007_1} we observe that, in all cases, POD$_{1}$ has very high percentages. Looking at the largest sample (sample A) the probability to predict a seeing weaker than 1" is 62.0 $\%$ (Table~\ref{tab:ct_seeing_val_2010_1}), while the probability to predict a seeing weaker than the first tertile is 48.1 $\%$ (Table~\ref{tab:ct_seeing_val_2010_2}). If we consider just the 2007 data (sample B), results are even better: we observe that the probability to predict a seeing weaker than 1" is 72.5 $\%$ (Tables~\ref{tab:ct_seeing_val_2007_1}) while the probability to predict a seeing weaker than the first tertile is 53.1 $\%$ (Tables~\ref{tab:ct_seeing_val_2007_2}). These values are well above the random cases of 33\%. We can say therefore that, in spite of the fact that there is obviously space for improvements, model performances in detecting the weakest values are therefore already very good, particularly in case in which we relax the threshold from the first tertile to 1". We analyzed separately the year 2007 because it gave some better results with respect to the total sample (2007/2010/2011). We think this might be due to the fact that the calibration sample belongs to the 2007 (also if the calibration sample is totally independent from the validation sample). This seems to indicate that there is space in the future to further improve the results with more sophisticated calibrations. The fact that POD$_{2}$ and POD$_{3}$ provide worse performances is justified by the fact that, in the scattering plot (Figure~\ref{fig:clouds_seeing_mnh_dimm}), the points show a tendency in increasing the dispersion for large values of the seeing. On the other side this is not necessarily a problem of the model (or only of the model) as we will see in Section \ref{cal_see}.

\subsubsection{Calibration sample}
\label{cal_see}

We report here the results for the calibration sample (20 nights sample). The interest of this analysis is that, for this sample, we can compare the performance of the model versus an instrument taken as a reference with the dispersion between two different instruments. This exercise provide us a very important reference on the real uncertainty with which we are at present able to estimate the seeing with measurements. This is therefore the ultimate limit over which a model can not goes. 

In Fig.\ref{fig:clouds_seeing_mnh_dimm_gs_20} are shown the scattering plots of the total seeing between GS and DIMM (left) and between the Meso-Nh and GS (center) and Meso-Nh and DIMM (right). We observe that the $\sigma$ value between the two instruments is 0.30" while those of the model with respect to the instruments is 0.34" and 0.36" therefore substantially comparable.
Tables~\ref{tab:ct_seeing_cal_gs_dimm_1} and \ref{tab:ct_seeing_cal_gs_dimm_2} report the contingency tables of the 2 instruments (GS and DIMM) in which we take, as a reference, one or the other instrument. Table~\ref{tab:ct_seeing_cal_gs_mnh} reports the contingency table between Meso-NH and the GS for the total seeing. Table~\ref{tab:ct_seeing_cal_dimm_mnh}  reports the contingency table between Meso-NH
and the DIMM for the total seeing. If we look at POD$_{1}$ i.e. the probability of the model to forecast a seeing weaker than the first tertile, this is 82.1 $\%$ (Table~\ref{tab:ct_seeing_cal_gs_mnh}) or 88.1 $\%$ (Table~\ref{tab:ct_seeing_cal_dimm_mnh}) depending if we consider the GS or the DIMM as a reference. These results are even better than those obtained comparing the two different instruments. Indeed the POD$_{1}$ i.e. the probability to measure a seeing weaker than the first tertile is 64.3 $\%$ (Table~\ref{tab:ct_seeing_cal_gs_dimm_1}) or 71.4 $\%$ (Table~\ref{tab:ct_seeing_cal_gs_dimm_2}) depending on the reference (GS or the DIMM). This tells us that the model fails on a percentage that is not larger than the intrinsic uncertainty with which we can estimate the turbulence.   

Besides that, to have informations on the trends we calculated the temporal evolutions. Fig.\ref{fig:temp_evol_J1}-Fig.\ref{fig:temp_evol_J3} show the temporal evolution of the turbulence integrated on the whole atmosphere (see Eq.\ref{eq:j}) as estimated by the Astro-Meso-Nh model and as measured by the GS and the DIMM in all the 20 nights of the PAR2007 campaign.
\begin{equation}
J = \int_{0}^{\infty }C_{N}^{2}(h)dh
\label{eq:j}
\end{equation} 

It is possible to appreciate how the model, during all the nights, reconstructs in a very satisfactory way not only the quantitative estimate of J but also the general trend (the morphology of the trends is well reconstructed). It is also interesting to observe that is extremely useful to run simultaneous different instruments because in some cases, the model appears more in agreement with one instrument than the other. Therefore, the fact to have more than one instrument permits us to better appreciate the real model performances. In other words, the intrinsic accuracy with which we can at present quantify the optical turbulence with an instrument is much lower than that achievable for a classical atmospheric parameters (such as temperature or wind) therefore the use of multi-references is extremely important in the analysis. 

%
%
\subsection{Wavefront coherence time - $\tau_{0}$}
\label{tau}

In the case of $\tau_{0}$ we could compare the Astro-Meso-Nh estimates with measurements coming from the DIMM and the MASS. When the model is compared to the DIMM we treated the whole 89 nights (2010, 2011 and 2007). When it is compared to MASS the analysis is limited to 48 nights because unfortunately there were no MASS measurements on 2007. For $\tau_{0}$ we could perform an analysis of model performances on summer as well as on winter time because the model calibration did not show particular limitations for this parameter.

\subsubsection{Validation sample}
\label{val_tau}

\begin{figure*}
\begin{center}
\begin{tabular}{c}
\includegraphics[width=0.7\textwidth]{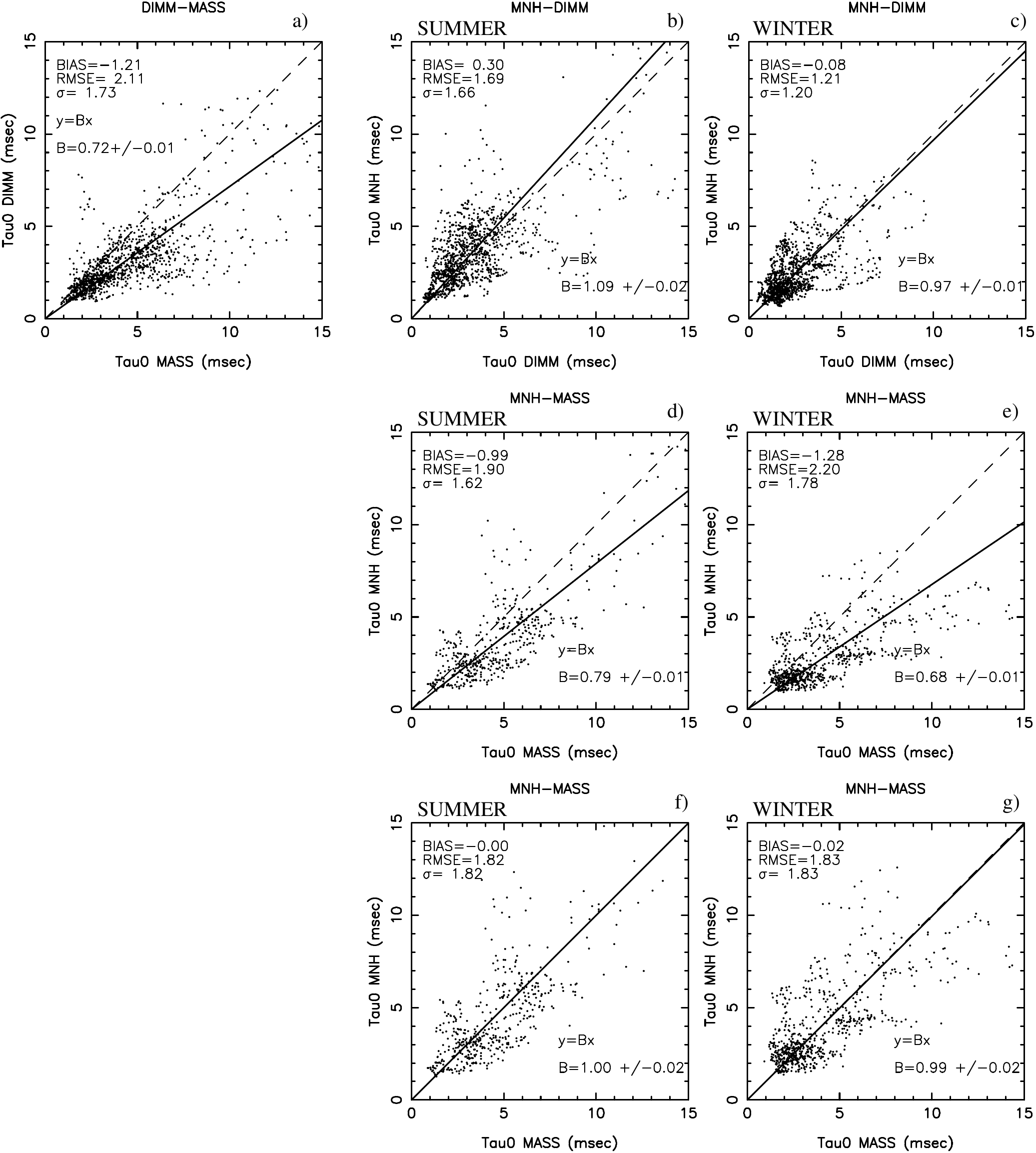}\\
\end{tabular}
\end{center}
\caption{\label{fig:cloud_plots_tauO_val} Scattered plots (validation sample) of the wavefront coherence time $\tau_{0}$ between (a) MASS and DIMM - 48 nights in 2010/1011 (there are no MASS
measurements in 2007); (b) Meso-NH and DIMM in summer - 42 nights in 2010/2011; (c) Meso-NH and DIMM in winter - 47 nights in 2010/2011; (d) Meso-NH and MASS in summer - 21 nights in 2010/2011; (e) Meso-NH and MASS in winter - 27 nights in 2010/2011; (f) the same as (d) but with the MASS corrected using the coefficient
of the regression line from (d); and (g) the same as (e) but with the MASS corrected using
the coefficient of the regression line from (e). For the model performances the key panels to take into account are (b), (c), (f) and (g). See extended discussion in the text.}
\end{figure*}

All the scattered plots for the wavefront coherence time $\tauO$ are summarized in Figure~\ref{fig:cloud_plots_tauO_val}. As can be seen in Fig.\ref{fig:cloud_plots_tauO_val}-(a) there is a not negligible bias between the DIMM and MASS measurements. This indicates that one or both instruments still have some problems to be arranged for $\tau_{0}$. We do not want to enter in a deep discussion on this issue because it is visibly not the main topic of the paper. What is important is that, assuming that the problem on the specific instrument is identified, this bias between the model and the instrument can be corrected by a multiplicative coefficient. The bias appears indeed visibly multiplicative. Even in case we can establish that  the MASS is the correct one and the DIMM is wrong, 
a multiplicative coefficient which corrects the regression line according to Fig.\ref{fig:cloud_plots_tauO_val}-(f) and (g) can be implemented.
The panels to be retained are therefore panels (b) and (c) if DIMM considered as reliable and panels (f) and (g) is MASS is considered reliable. 
For what concerns the scattering plots, we note that the values of $\sigma$ obtained between DIMM and MASS and between Meso-Nh and MASS or Meso-Nh and DIMM are comparable. In some cases the $\sigma$ of Meso-Nh with respect to the instrument (for example the $\sigma$ of Meso-Nh and DIMM in summer and winter is respectively of 1.66 ms and 1.20 ms) is even smaller than that obtained considering the two instrument ($\sigma$=1.73 ms). 

From a different perspective, Tables~\ref{tab:ct_mass_dimm_tauO_val} and \ref{tab:ct_dimm_mass_tauO_val} report the contingency tables between MASS and DIMM for $\tauO$ on the whole sample of 48 nights of the 53 simulated nights of the period 2010-2011 in which we had simultaneous measurements from DIMM and MASS. In the first table the DIMM is taken as a reference, in the second table the MASS is taken as a reference. The very different values of POD$_{1}$ and POD$_{3}$ in both tables just reflects the problem we discussed for the scattering plot. MASS and DIMM have a visible not negligible bias that, however, in phase of analysis of model performances, can be corrected with a multiplicative coefficient (See panels (f) and (g) of Fig.\ref{fig:cloud_plots_tauO_val}).
Tables~\ref{tab:ct_dimm_mnh_summer_tauO_val} is the contingency table between Meso-NH and DIMM for the 42 nights (over the total of 89 nights of the validation sample) related to summer.
Tables~\ref{tab:ct_dimm_mnh_winter_tauO_val} is the contingency table between Meso-NH and DIMM for the 47 nights (over the total of 89 nights of the validation sample) related to winter.
Tables~\ref{tab:ct_mass_mnh_summer_tauO_val} is the contingency table between Meso-NH and MASS for the 21 nights (over the total of 48 nights of the validation sample) related to summer.
Tables~\ref{tab:ct_mass_mnh_winter_tauO_val} is the contingency table between Meso-NH and MASS for the 27 nights (over the total of 48 nights of the validation sample) related to winter.  Table~\ref{tab:ct_mass_mnh_summer_tauO_val} and Table~\ref{tab:ct_mass_mnh_winter_tauO_val}  have been calculated using obviously the data corrected as in Fig.\ref{fig:cloud_plots_tauO_val}-(f) and (g). 

As we did for the seeing we focused our attention on the most crucial POD$_{i}$ for the $\tau_{0}$. Differently from the seeing, the most critical and interesting POD for $\tau_{0}$ is POD$_{3}$ that is the probability to forecast the $\tau_{0}$ that is larger than the second tertile of the cumulative distribution. When the $\tau_{0}$ is large, an AO system can be run at low temporal frequency. This not only allows for better AO performances but it facilitates the achievement of scientific programs that could not be completed otherwise.  Looking at contingency tables concerning the model (Table~\ref{tab:ct_dimm_mnh_summer_tauO_val}, Table~\ref{tab:ct_dimm_mnh_winter_tauO_val}, Table~\ref{tab:ct_mass_mnh_summer_tauO_val} and Table~\ref{tab:ct_mass_mnh_winter_tauO_val}) we observe that we obtain, in all cases, very good values of POD$_{3}$. POD$_{3}$ = 70.7 $\%$ for Meso-Nh-DIMM (summer time, Table~\ref{tab:ct_dimm_mnh_summer_tauO_val}) and 65.6 $\%$ for Meso-Nh-DIMM (winter time, Table~\ref{tab:ct_dimm_mnh_winter_tauO_val}). POD$_{3}$ = 71.9 $\%$ for Meso-Nh-MASS (summer time, Table~\ref{tab:ct_mass_mnh_summer_tauO_val}) and 78.2 $\%$ for Meso-Nh-MASS (winter time, Table~\ref{tab:ct_mass_mnh_winter_tauO_val}). Also the values of POD$_{1}$ that tell us the probability to detect a $\tau_{0}$ weaker than the first tertile is very satisfactory included in the range [48-69] $\%$ depending of the instrument we take as a reference. 

For what concerns the discrepancy/bias between MASS and DIMM at present we can say that \cite{masciadri2014} proved that $\tau_{0}$ from MASS appears reliable. Only when values are large (larger than 5 ms) the instrument shows a tendency in slightly overestimate $\tau_{0}$. On the the other side the DIMM is not a vertical profiler and for this reason is not an instrument particularly suitable to quantify $\tau_{0}$, a parameter strongly dependent on the vertical stratification of $\CN2$ and wind speed. The method used to retrieve $\tau_{0}$ from the DIMM has been proposed by \cite{sarazin2001} but the same authors affirm that the solution is just based on approximations and has to be considered with the suitable precautions.

%
%
\subsection{Isoplanatic angle - $\theta_{0}$}
\label{theta}

In the case of $\theta_{0}$ we could compare the Astro-Meso-Nh estimates with measurements coming from the DIMM and the MASS. However, as we have done for the seeing, we present only comparisons model/measurements obtained in summer time because they are the only useful results. This is because $\theta_{0}$ require a separate calibration in summer and winter time as well as the seeing $\varepsilon$.

\subsubsection{Validation sample}
\label{val_theta}
 
\begin{figure}
\begin{center}
\begin{adjustbox}{max width=\columnwidth}
\begin{tabular}{c}
\includegraphics[width=0.5\textwidth]{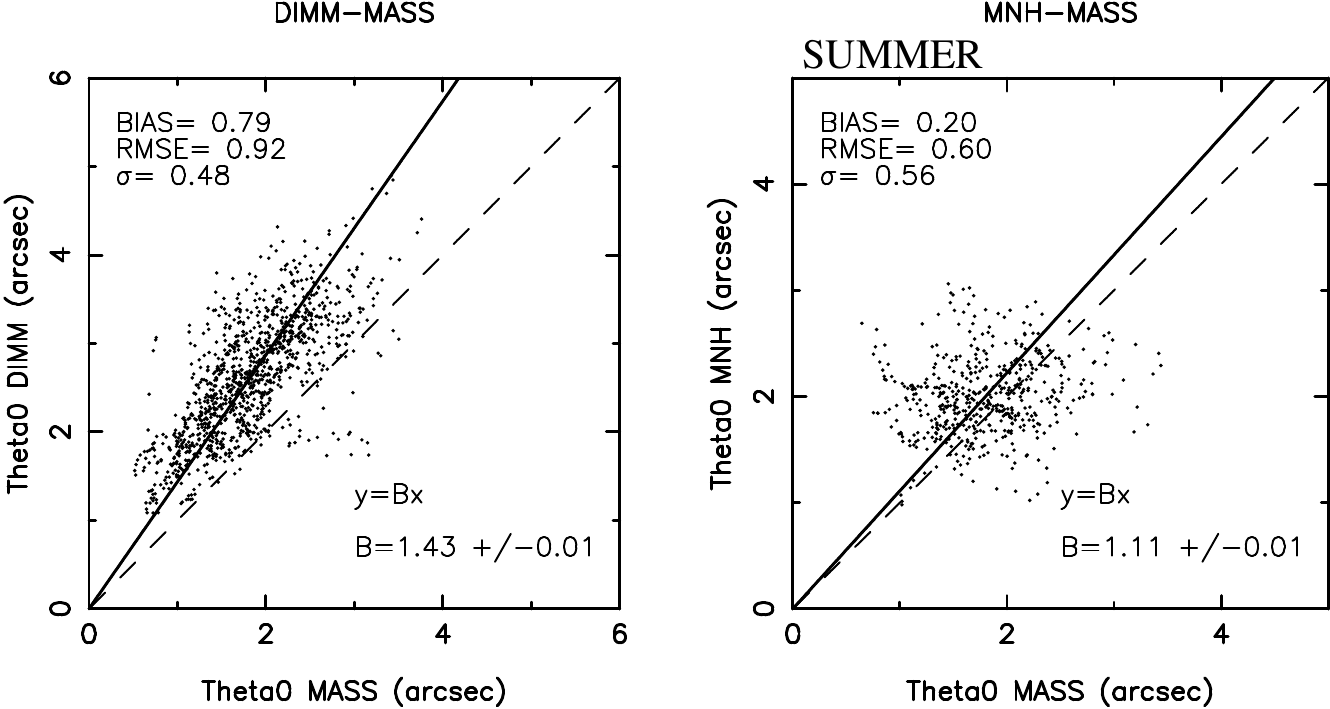}\\
\end{tabular}
\end{adjustbox}
\end{center}
\caption{\label{fig:clouds_thetaO_val} Scattered plots of the isoplanatic angle $\theta_{0}$ between DIMM and MASS (left) and between MASS and Meso-NH (right).}
\end{figure}

Figure~\ref{fig:clouds_thetaO_val} shows the scattered plots of $\thetaO$ between DIMM and MASS on the sub-sample of 44 nights over the 53 nights in 2010/2011 in which there were simultaneous measurements (left) and between MASS and the model for the summer sample corresponding to 21 nights over the total sample of 48 nights in 2010/2011 (right). 

Table~\ref{tab:ct_dimm_mass_thetaO_2010} and \ref{tab:ct_mass_dimm_thetaO_2010} are the contingency tables for \thetaO\ between MASS and DIMM for the 44 nights 
in 2010/2011. Table~\ref{tab:ct_mass_mnh_thetaO_2010_summer} is the contingency table between Meso-NH and MASS, during summer (21 nights over 48 nights) for \thetaO. 
We observe that, in a similar way than for $\tau_{0}$, the MASS and DIMM distribution of points in Fig.\ref{fig:clouds_thetaO_val} presents a not negligible bias. However, we know that a study from \cite{masciadri2014} proved that the $\theta_{0}$ from MASS is well correlated to that of GS (at least on a sample of 14 nights) and $\theta_{0}$ estimates from MASS can be considered reliable. No biases have been observed between MASS and GS. We think therefore that, for the same arguments/reasons presented in the case of $\tau_{0}$, we can reasonably think that the problems are for $\theta_{0}$ in DIMM measurements (and not in MASS). The method proposed by \cite{sarazin2001} to estimate $\theta_{0}$ and $\tau_{0}$ with a DIMM is based on an approximation and can not be considered very accurate (as said by the same authors). We concentrated here therefore only on a comparison model/MASS measurements.
Observing the scattering plots of Fig.\ref{fig:clouds_thetaO_val} we note that the two figures have a comparable value for $\sigma$ (0.48" vs. 0.56") even if measurements are better correlated among them than Astro-Meso-Nh model versus MASS\footnote{We remind that $\sigma$ is not affected by the bias.}. The cloud of points in the panel on the left is indeed better elongated along the regression line than panel on the right. 

\begin{figure}
\begin{center}
\begin{adjustbox}{max width=\columnwidth}
\begin{tabular}{c}
\includegraphics[width=0.5\textwidth]{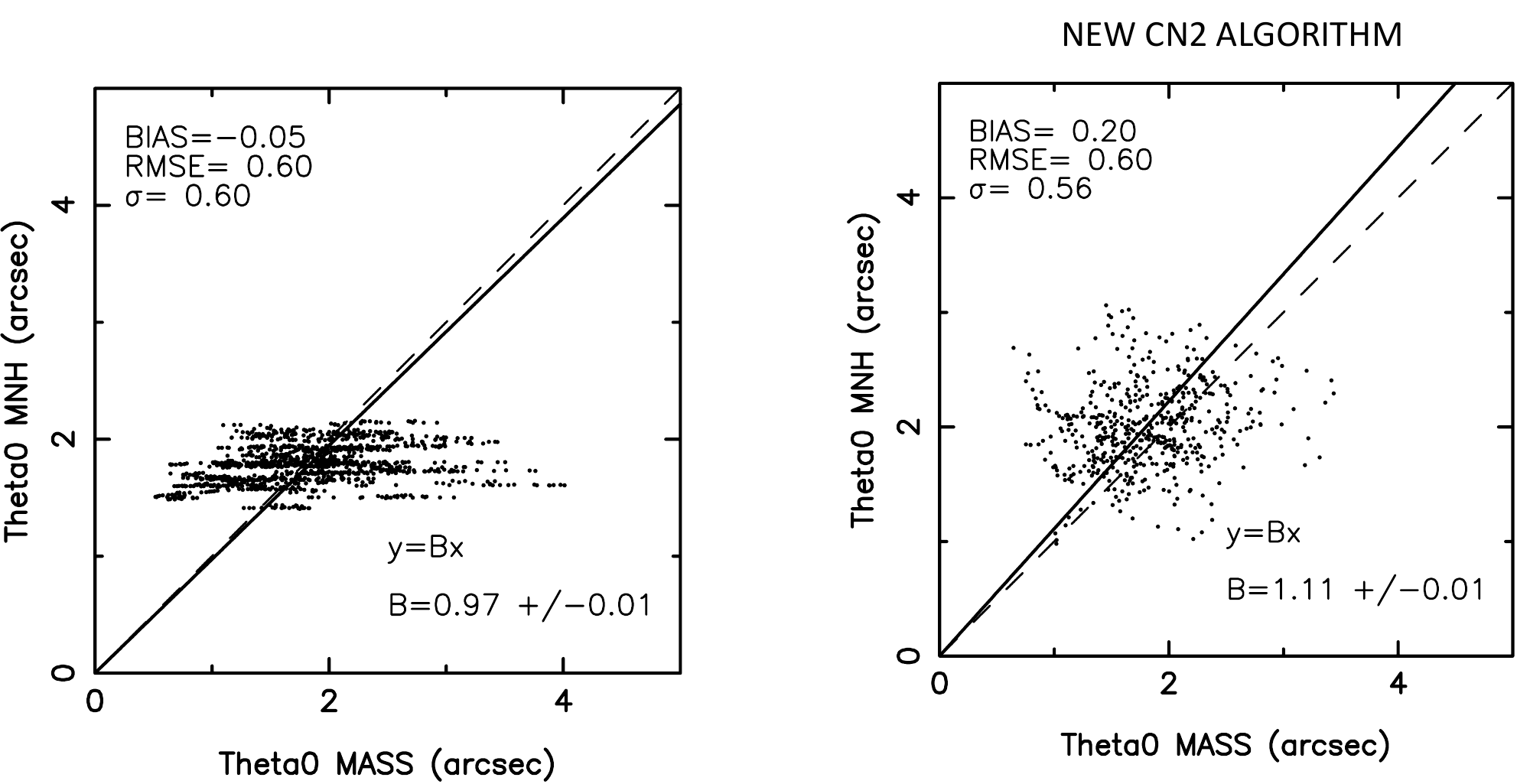}\\
\end{tabular}
\end{adjustbox}
\end{center}
\caption{\label{fig:theta0_before_after} Scattered plots of the isoplanatic angle $\theta_{0}$ between MASS and Meso-NH before (left) and after (right) the implementation of the new $\CN2$ algorithm.}
\end{figure}

On the other side it is possible to observe that the model has visibly improved its performances for the reconstruction of $\theta_{0}$ using the new algorithm of the $\CN2$ (Eq.\ref{eq:cn2_mod_1}). Fig.\ref{fig:theta0_before_after} shows indeed the distribution before and after the modification of the algorithm. It is evident that before the modification of the algorithm the model had serious problems in well reconstructing the spatio-temporal variability of the turbulence in the high part of the atmosphere during the night. $\theta_{0}$ is indeed mainly driven by the turbulence in the free atmosphere. $\theta_{0}$ covers now almost on the same interval of values observed with the MASS (typically [1,3] arcsec)\footnote{We remind that measurements and model outputs are both treated with a moving average of 1 h. Raw variability is larger in both measurements and simulations}. Even if there is still space for improvements in the algorithm, we have now comparable dispersion of model values with respect to measurements. 

If we analyze the contingency tables and the values of POD$_{i}$ (Table~\ref{tab:ct_mass_mnh_thetaO_2010_summer} ) we observe that they are some how weaker than those of the seeing $\varepsilon$ and $\tau_{0}$. However, the most interesting POD$_{i}$  from an astronomical point of view for $\theta_{0}$ is  POD$_{3}$ that is the probability that the model detects a $\theta_{0}$ larger than the second tertile. POD$_{3}$ permits us to identify temporal windows with large $\theta_{0}$ and those intervals are favorable to ground-based observations supported by GLAO, MCAO and WFAO systems. Our system reveals therefore useful in particular to the Adaptive Optics Facility (AOF - \citealt{arsenault2014,madec2016,kuntschner2012}) the ESO AO system for the unit UT4 of the VLT that will use an adaptive secondary and it is composed by the two modules GALACSI and GRAAL. Our estimate of POD$_{3}$ is of the order of 58.8 $\%$ therefore quite satisfactory and almost the double of the 33\% of the random case. POD$_{1}$ is still weak and this tells us that there is still space for improvements (even if POD$_{1}$ is definitely less critical from the point of view of the Service Mode). 

%
%
\subsection{Discussion}
\label{discussion}

We report here a few consideration to complete the analyses presented.

\begin{enumerate}

\item The use of POD$_{i}$ much more than PC (that is associated to model performances as a whole) can support us to quantify a concrete score of success calculated at posteriori on a rich statistical sample of estimates. We observed that if we take into accounts all the POD$_{i}$ (POD$_{1}$, POD$_{2}$ and POD$_{3}$) we find sometimes very satisfactory results, other times results are more modest. If we limit us to the important POD$_{i}$ for each astroclimatic parameter i.e. POD$_{1}$ for the seeing $\varepsilon$ and POD$_{3}$ for $\theta_{0}$ and $\tau_{0}$, results are always very satisfactory and, in all cases, the relative PODs are much better than 33$\%$. In many cases they are two times better and in some cases even more than this. Our results indicates therefore that the impact on the Service Mode is concrete. In the previous session we quantified that. 

What about the impact of our forecast method with respect to other strategies ? We cite here just a couple of alternative approaches. The first method is to have no system at all for the forecast that corresponds to the random case. In that case all the POD$_{i}$ are equal to 33$\%$. Our system is visibly better than that. To appreciate the difference we would obtain using as a forecast method the climatological median (of whatever astroclimatic parameter) the reader can have a look to the contingency table shown in Table~\ref{tab:ct_climatology}. The method of the climatological median provides a probability to detect the parameter we intend to study of 100 $\%$ for values between the first and second tertile but a probability of 0 $\%$ for values smaller or larger than the first and second tertiles. These latter intervals are, however, the most interesting one from an astronomical point of view. The real critical temporal windows in which we are interested on are those characterized by a very weak seeing during which to select most challenging programs for example for the search and characterization of exo-planets or, in alternative, those temporal windows with a very large $\tau_{0}$ that can permit us to use the AO system in their best configuration with the best performances permitting the best sky coverage or, in alternative, the temporal windows with the largest $\theta_{0}$ that guarantee us very weak turbulence conditions in the free atmosphere. Those are the best conditions to use WFAO systems on very large field of view for science cases such as, for example, resolved stellar population studies and photometry in globular clusters \citep{fiorentino2016}. If we trust in the climatology median we will be more efficient in detecting the right time in which the parameter we are studying is close to the median climatology but we will never identify those precious temporal windows close to the extremes values. The Astro-Meso-Nh model is visibly much more effective in this respect. Another methods might be envisaged such as the prediction for persistence that assumes that the value of a parameter remains equal to itself. In reality one might imagine many different versions of such a method: (a) the value at the beginning of the night remains the same for the whole night, (b) the seeing remains the same on intervals of time inside the duration of the night, (c) others. We planned to study these cases in the near-future when we will have a larger statistics and when a more clear strategy of the Service Mode will be defined\footnote{This will be discussed with the Science Operation Division of ESO and the ESO MOSE Board}. It is worth to say that, in general, an atmospheric model can be more effective when we want to put in evidence some changes with respect to the previous conditions. Under conditions of perfect stability an atmospheric model is less useful. \\
 
\begin{table}
\begin{center}
\begin{adjustbox}{max width=\columnwidth}
\begin{tabular}{ccccc}
\multicolumn{2}{c}{Climatology} & \multicolumn{3}{c}{\bf OBS}\\
\multicolumn{2}{c}{whatever N. of nights} &  par. $<$ X1  &  X1 $< $ par. $<$ X2   &  par.$ >$ X2 \\
\hline
\multirow{7}{*}{\rotatebox{90}{\bf Median Value }} & & &\\
& par. $<$ X1     & 0  & 0 & 0  \\
 &      &  &  & \\
 &  X1 $<$ par. $< $X2 & 33 & 33 & 33 \\
 &      &  &  & \\
 & par. $>$ X2                    & 0 & 0 & 0  \\
 &      &  &  & \\
\hline
\multicolumn{5}{l}{$PC$=33\%; $EBD$=0\%} \\
\multicolumn{5}{l}{$POD_1$=0\%; $POD_2$=100\%; $POD_3$=0\%} \\
\end{tabular}
\end{adjustbox}
\caption{\label{tab:ct_climatology} Typical contingency table using the climatological median value (as N is undefined, we display the
percent only).}
\end{center}
\end{table}

\item An interesting parameter that can be estimated by the Astro-meso-Nh model is the fraction of the turbulent energetic budget contained in the first 600 m with respect to the turbulent energetic budget included in the whole 20 km (J$_{BL}$/J$_{TOT}$). Fig.\ref{fig:frac_energ} shows the temporal evolution of J$_{BL}$/J$_{TOT}$ during the whole night of PAR2007 campaign as measured by the GS and as reconstructed by the model. In the figure it is visible that the model well reconstructs the growing of J$_{BL}$/J$_{TOT}$ during the night. This figure of merit is extremely useful for the observations to be done with whatever wide field adaptive optics system (MCAO, GLAO, MOAO, LTAO). In the context the MOSE project we performed this analysis in perspective to an application to AOF at the VLT (\citealt{madec2016,kuntschner2012}). \\

\begin{figure} 
\begin{center}
\begin{adjustbox}{max width=\columnwidth}
\begin{tabular}{c}
\includegraphics[width=0.7\textwidth]{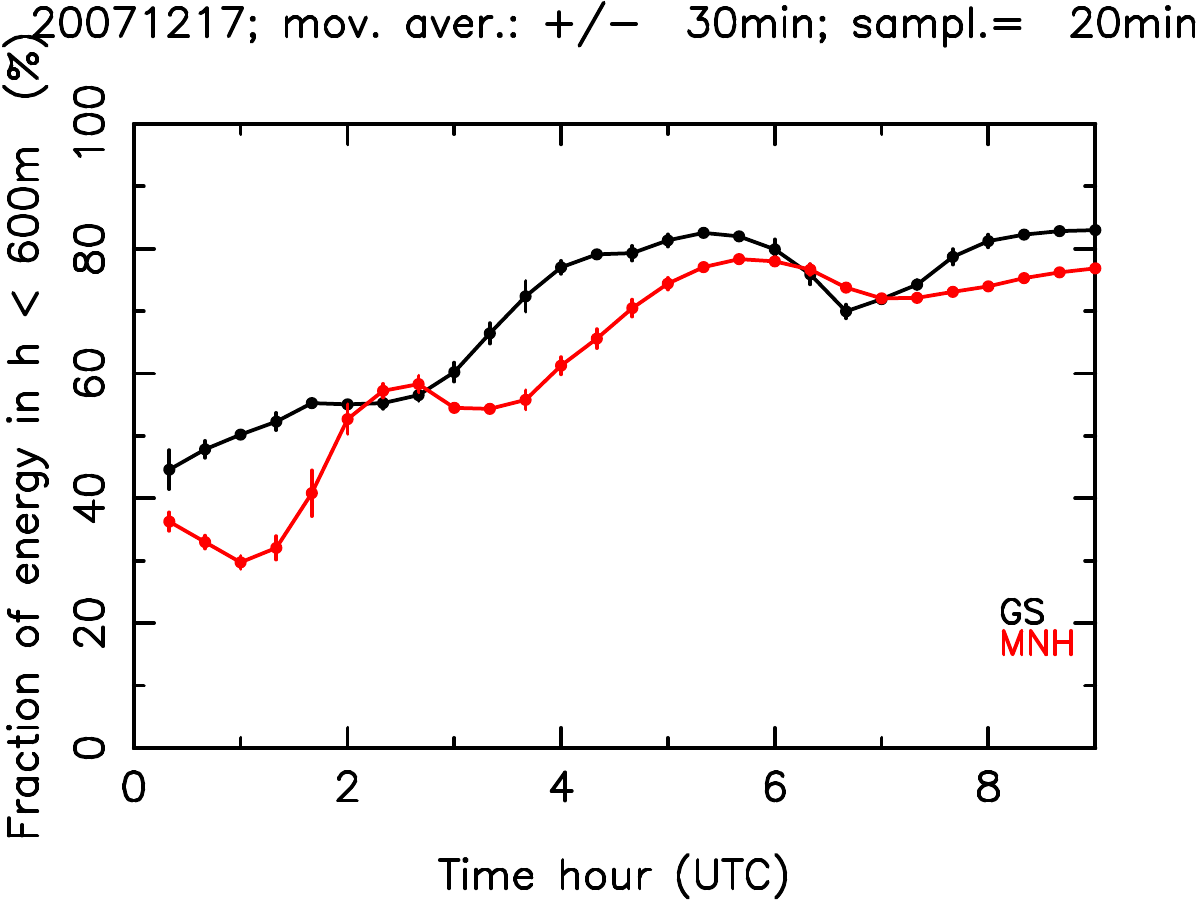}
\end{tabular}
\end{adjustbox}
\end{center}
\caption{\label{fig:frac_energ} Temporal evolution of the fraction of turbulent energy in the first 600 m with respect to the total turbulence energy on the whole 20 km as measured by the Generalized SCIDAR and as estimated by the Astro-Meso-Nh model during one night of the PAR2007 site testing campaign.}
\end{figure} 

\item In general, model performances tends to decreases if it is initialized and forced with 'forecasts' instead of 'analyses' coming from the GCM. The level of the deterioration depends in general on the forecast delay. However we could prove \citep{masciadri2015} that, for all the atmospheric parameters and for the forecast delay we selected for the operational configuration, the PODs obtained with the ECMWF forecasts are weakly worse than those obtained with the ECMWF analyses, and sometime they present the same value. The impact seems therefore weak and therefore comfortable. Due to the fact that the $\CN2$ depends on the atmospheric parameters, we deduce that the same conclusion is valid for the $\CN2$.  \\

\item In the analysis we presented it appears evident that the necessity of rich statistical sample of measurements taken with different instruments is fundamental to progress in this research. Also we observed that the uncertainty in quantifying the astroclimatic parameters with measurements is not negligible in some cases. We observed for example that the $\sigma$ between GS and DIMM (PAR2007 campaign) was of 0.30". It has therefore no sense to pretend a better result than this for a model. \\

\item Results obtained in this study indicates that it is very important to be able to run routinely and simultaneously above all observatories as many as possible different instruments based on independent principles.  

\end{enumerate}
%
%
\section{High vertical resolution C$_{N}^{2}$ profiles}
\label{cn2_hvr} 

All results and analyses presented so far implied the use of a model configuration having 62 vertical levels covering the whole atmosphere (roughly 20 km above the ground). The vertical resolution of the Astro-Meso-Nh model is of the order of 600 m above 3.5 km above the ground (i.e. in the central and high part of the atmosphere). It is evident that, the higher the model vertical levels number, the higher is the computing effort. A $\CN2$ with a resolution of $\sim$ 600 n the central part of the atmosphere can be considered a good baseline because it is enough for most of applications of operations in Service Mode and in AO context. Also we note that the model configuration we selected is suitable to perform nightly automatic operational forecasts using a relative cheap hardware, in other words to implement an operational forecast system that does not require super clusters. 
However, for most sophisticated AO systems, typically the Wide Field Adaptive Optics (WFAO) i.e. GLAO, MCAO, LTAO, MOAO it is extremely important to know $\CN2$ profiles with vertical resolutions as high as 150~m \citep{costille2012}. The tomographic error is, indeed, strongly dependent from the $\CN2$ profiles and a lack of knowledge of the vertical stratification with a sufficiently high vertical resolution can induce to large tomographic errors.
These AO systems are currently tested and implemented on the present generation facilities and they are planned to be implemented on the ELTs in the next decades. For example at the E-ELT the near-infrared spectrograph HARMONI \citep{thatte2016} will work with a LTAO system, MAORY \citep{diolaiti2016} is a MCAO system that will feed the imager MICADO \citep{davies2016} and the multi-object spectrograph MOSAIC \citep{hammer2016} will be feed by a MOAO system. The wider is the field of view, the higher is the resolution with which one has to know the vertical stratification of the turbulence in advance. The most demanding AO system is the MOAO that requires the highest resolution (see a summary in \citealt{masciadri2013c}). 

\begin{figure*}
\begin{center}
\begin{tabular}{c}
\includegraphics[width=0.7\textwidth]{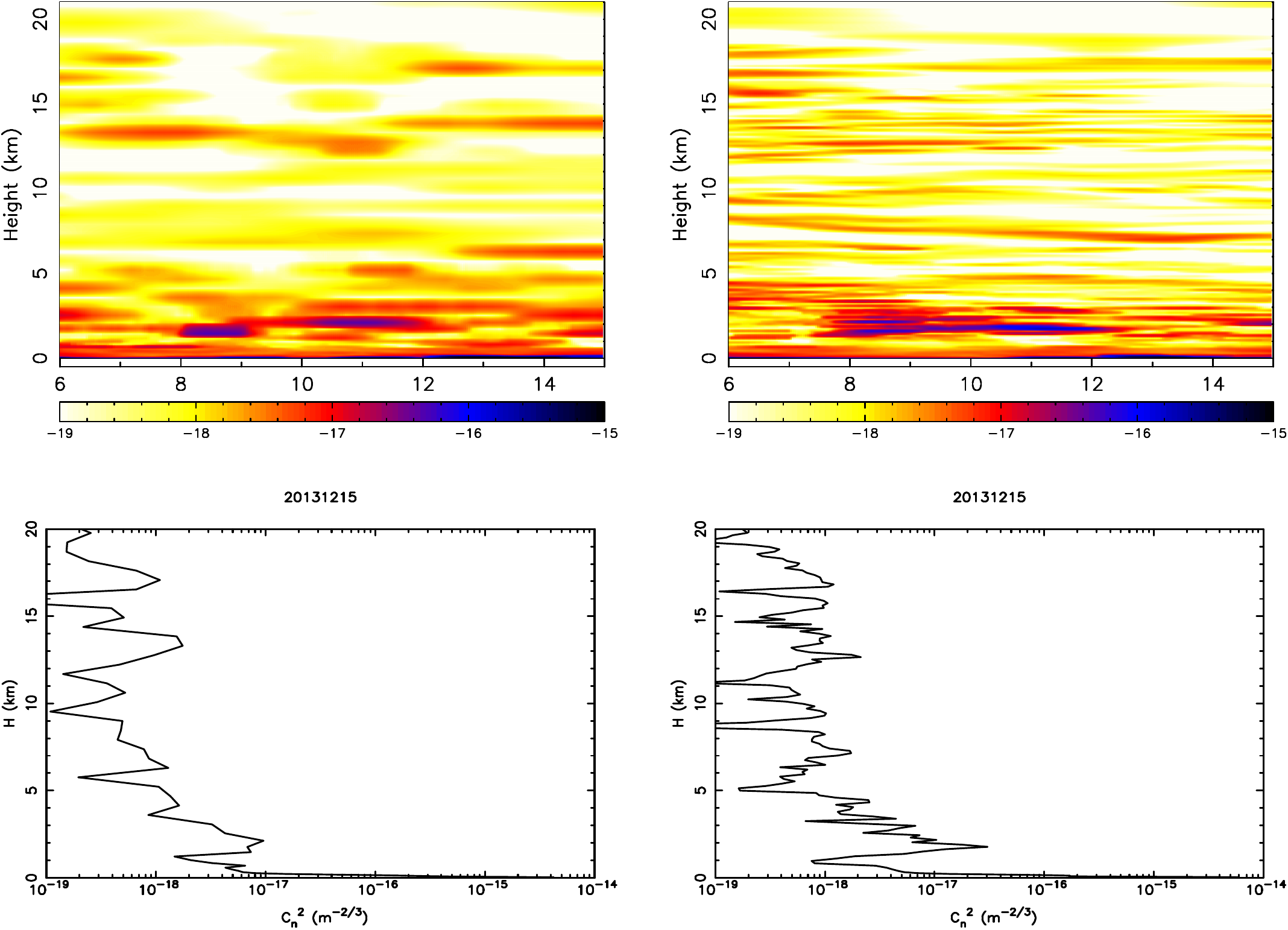}\\
\end{tabular}
\end{center}
\caption{\label{fig:cn2_hvr_tot} Top: Temporal evolution of the vertical $\CN2$ profiles simulated by the Astro-Meso-Nh model with different model vertical resolutions (62 levels (left), 173 levels (right)) related to the night 15/12/2013. Bottom: corresponding average $\CN2$ profile related to the $\CN2$ temporal evolution on the top. }
\end{figure*}

For this reason we therefore put a great effort in trying to improve the vertical resolution of the $\CN2$ in perspective of an implementation in an operational forecast system. In Fig.\ref{fig:cn2_hvr_tot} we present the fundamental results we achieved. On the top we have the temporal evolution of the $\CN2$ along a night obtained by the Astro-Meso-Nh model having 62 vertical levels (left) and 173 levels (right) with the new $\CN2$ algorithm. On the bottom is reported the average $\CN2$ of the night related to the respective simulations. It is well visible that the simulations done with 173 vertical levels provide a $\CN2$ with a much larger number of turbulent layers. Also the turbulent layer are much thinner than those resolved by the 62 levels configurations. This result was possible thanks to the new $\CN2$ algorithm presented in this paper (see Section \ref{eq:cn2_mod_1}) that permits us to better represent the spatio-temporal variability of the vertical stratification of the OT in the 20 km above the ground.
The $\CN2$ shown in Fig.\ref{fig:cn2_hvr_tot} proves the ability of our technique in reconstructing a vertical stratification of the optical turbulence with a resolution that opens crucial and exciting new perspectives in the field of the WFAO. We calculated, indeed, that we should be able to use this model configuration also for an operational model configuration. Since the model calibration depends on its configuration we planned to calibrate the model with this configuration in the near-future. In the meanwhile, this $\CN2$ profile has been used for a preliminary study aiming to testing the performances of HARMONI \citep{neichel2016} and for a more extended study aiming to address the ultimate limitations of the WFAO technique\footnote{The study has been presented by Thierry Fusco at al. at SPIE Conference, Adaptive Optics Systems V, 2016, 9909-273.}. 

%
%
\section{Conclusions}
\label{concl}

In this paper we presented the main results we obtained on the model performances in forecasting the optical turbulence above Cerro Paranal in the context of the MOSE project. 
We analyzed the three main integrated astro-climatic parameters as originally planned: seeing ($\varepsilon$), isoplanatic angle ($\theta_{0}$) and wavefront coherence time ($\tau_{0}$) that fundamental in the AO optimization and planning of ground-based observations. The Astro-Meso-Nh model has been firstly calibrated using measurements taken with a Generalized SCIDAR. The model performances have been quantified using measurements coming from samples of nights completely independent from the calibration one. The results we obtained on the validation samples are summarized in a nutshell:\\

\begin{enumerate}

\item The wavefront coherence time ($\tau_{0}$) is the parameter that, at present time, is predicted with the best score of success. All the POD$_{i}$ (i = 1, 2 and 3) are satisfactory and comparable to those obtained among different instruments, in some cases even better. From an astronomical point of view, the most interesting POD$_{i}$ for $\tau_{0}$ is POD$_{3}$ that is the probability to detect a $\tau_{0}$ larger than the second tertile of the cumulative distribution of measurements. We are therefore interested in the temporal windows in which the $\tau_{0}$ is large.
POD$_{3}$ on the validation sample is included in the range [72$\%$ - 78$\%$] if we considered the most reliable instrument as a reference i.e. the MASS. The range includes the analysis done in different seasons: summer and winter. The samples we are dealing about include 42 nights (summer) and 47 nights (winter). POD$_{1}$ is included in the range [48$\%$ - 69$\%$] (slightly smaller than POD$_{3}$ but still very good) using as a reference the same instrument. Another not negligible positive element for $\tau_{0}$ is that this score of success has been obtained with just one model calibration  valid for the whole year (i.e. not a calibration for summer and one for winter) and we analyzed data in summer as well as in winter. It is highly probable that this score of success would further improve with a seasonal model calibration. 
The ability in detecting a large $\tau_{0}$ in advance is extremely useful for the following reasons. A large $\tau_{0}$ permits an AO system to run at low temporal frequency and to use faint stars as guide stars. A low temporal frequency permits indeed to better sample light coming from stars that, in presence of a smaller $\tau_{0}$, can not be used as guide stars. A large $\tau_{0}$ represents an advantage for the wide field AO because the number of available guide stars, in general, increases (i.e. the sky coverage increases) and because a higher Strehl Ratio (SR) is reachable and more ambitious scientific projects can be carried out. The advantages are also for the classical SCAO configuration (not only for WFAO) because it permits to observe fainter stars reaching a higher SR.\\

\item The total seeing ($\varepsilon$) revealed to be sensitive to the seasonal model calibration. This means that the model needs a twofold calibration: one for summer and  one for winter. We provided therefore in this study only the score of success for the summer time because the calibration sample belongs to this season. However the goal of this study was to prove the feasibility of the OT forecast therefore results we obtained fit perfectly with objectives. We will be able to use the measurements of the site testing campaign planned for Paranal in the near-future with a Stereo-SCIDAR \citep{osborn2016} to calibrate the model for the winter time. In the case of the seeing, the most interesting POD from a scientific point of view is POD$_{1}$. 
The probability to detect the $\varepsilon$ $<$ 1" is in the range [62$\%$ - 72$\%$]. The probability to detect the $\varepsilon$ $<$ of the first tertile is in the range [48$\%$ - 53$\%$]. 
The range of values in this case have the following meaning: the most conservative values (62$\%$ and 48$\%$ have been obtained with the sample of 42 nights (summer) in the 2010/2011 years. The most optimistic values have been obtained with the sub-sample of the 18 nights (summer) of 2007 (the same year in which the calibration sample belongs). This indicates that the identification of a sample of measurements used for the calibration is an important issue in the process. We envisage an as extended as possible sample of nights for that in the future. The instrument used as a reference is in this case the DIMM. 
The advantage in detecting in advance temporal windows with weak $\varepsilon$ is obvious. The best seeing conditions guarantee the highest SR of an AO system and the most challenging scientific programs such as the search and the characterization of extra-solar planets would definitely gain from this condition. Besides we note that the selection of the temporal windows with excellent seeing will be even more critical for the adaptive optics in the visible \citep{close2016} that is much more dependent on the conditions of the turbulence than the AO in the near-infrared. This new typology of adaptive optics represents the new frontier of ground-based observations. \\

\item The isoplanatic angle ($\theta_{0}$) revealed to be sensitive to the seasonal model calibration as is the case for the seeing too. We provide therefore in this study only the score of success of the model related to the summer time for the same reason described for the seeing. From a scientific point of view, for $\theta_{0}$ the most interesting POD$_{i}$ is POD$_{3}$ that is the probability to detect a $\theta_{0}$ larger than the second tertile. We obtained a very promising 58\% for POD$_{3}$ on the validation sample using, as a reference, the MASS. The sample we are dealing about includes 21 nights (summer) of (2010/2011). No measurements with MASS are available for 2007. For completeness we also note that, at present, the probability to detect an isoplanatic angle smaller than the first tertile (POD$_{1}$) is not sufficiently good and we plan to undertake further investigations to improve this score of success.  At the same time we highlight the fact that, from a scientific point of view, in the context of the Service Mode POD$_{3}$ is definitely more interesting than POD$_{1}$. We think therefore that, at present, our forecasting system is definitely able to provide useful informations for the Service Mode. 
A large value of the $\theta_{0}$ is particularly useful for whatever kind of wide field AO (GLAO, LTAO, MCAO, MOAO). Such a systems in general correct turbulence close to the ground and if we are able to detect temporal windows with not so much turbulence in the free atmosphere (large $\theta_{0}$), ww should be able to give a crucial input for the Service Mode. Observations supported by AO systems of extended and crowded fields such as clusters would definitely benefit from this condition if it would be known in advance. \\

\item The general commentary is that, as expected, model performances in reconstructing the astroclimatic parameters are not as good as in reconstructing the atmospheric parameters (treated in \citealt{masciadri2013, lascaux2013,lascaux2015}). However, if we look at the model performances in reconstructing values of $\varepsilon$, $\tau_{0}$ and $\theta_{0}$ in the most critical and interesting ranges of values for each of them, model performances are very good. Quantitive results indicate that the Astro-Meso-Nh model performances in forecasting the optical turbulence have definitely a fundamental impact on the Service Mode of top-class telescopes. \\

\item We proved to be able to forecast $\CN2$ profiles with a vertical resolution as high as 150~m. As discussed in the paper, such an achievement definitely opens interesting new perspectives for the development of the most sophisticated AO systems planned for the next generation ground-based telescopes. \\

\end{enumerate}

In terms of perspectives we intend to continue our studies/work on the $\CN2$ algorithm in order to improve the model performances. There is still space to improve the mixing length definition in stable regime and this is a research line that we will follow with attention. The access to new measurements of the stratification of the turbulence ($\CN2$) that are planned to be done above Paranal will be important for a calibration of the model on the whole year. A richer sample of measurements will be also useful to enlarge the typology of analysis. Our kind of research will definitely take advantage from the fact to have as many as possible different reliable instruments running simultaneously above a site because this permits to better constrain the model but also to better quantify the uncertainty with which we can estimate the turbulence. In perspective to an operational implementation we planned, in agreement with the ESO Science Operational division and the ESO MOSE Board, some more dedicated studies on the impact of the forecast on the Service Mode. 

\section*{Acknowledgments}
This study is co-funded by the ESO contract: E-SOW-ESO-245-0933 (MOSE Project).
We are very grateful to the ESO Board of MOSE (Marc Sarazin, Pierre-Yves Madec, Florian Madec and Harald Kuntschner) for their constant support to this study.  Part of measurements used in this study were part of the PAR2007 site testing campaign. A great part of simulations are run on the HPCF cluster of the European Centre for Medium Weather Forecasts (ECMWF) - Project SPITFOT. 







\clearpage
\appendix
%
%
\section{$\CN2$ algorithm}
\label{annex_a}

A few differences have to be signalled with respect to the $\CN2$ algorithm of \citep{masciadri1999}. The main goal of this paper is to show the model performances and we expressly by-pass extended analyses on the testing of different algorithms that lead us in using this algorithm for this analysis. For clarity, we report here a few informations that in the context of this paper play a secondary role but that just clarify differences in the $\CN2$ algorithm with respect to \citep{masciadri1999}. The mixing length in this paper (Eq.\ref{eq:lm}) is different from that of \citet{bougeault1989} reported in \citet{masciadri1999}. This is due to the fact that, more recently, the Meso-Nh code assumed this new formulation\footnote{See Scientific Documentation Meso-Nh at \url{http://mesonh.aero.obs-mip.fr/mesonh52}} for mainly two reasons ({\it Valery Masson private communication}): (1) the new expression better fits with the diagnostic of the surface layer in convective regime in LES configuration, (2) the new formulation permits to have a linear dependency in the limit of the surface layer where the the mixing length of the turbulence is proportional to the distance from the ground. Point (1) is mainly irrelevant for our application.

We point out also a difference in Eq.\ref{eq:cn2_2} with respect to that of a precedent used \citep{masciadri1999}. In this paper we used, indeed, the factor:\\
\begin{equation}
\left ( \frac{80\cdot 10^{-6}\cdot P(z)}{T(z)\cdot \theta (z)} \right )^{2}
\label{ann1:eq1}
\end{equation}
instead of:\\
\begin{equation}
\left ( \frac{80\cdot 10^{-6}\cdot P(z)}{T(z)^{2}} \right )^{2}
\label{ann1:eq2}
\end{equation}

Eq.\ref{ann1:eq2} (also called Gladstone's relation) is found in the literature \citep{roddier1981,vernin2002} and assumes that the atmosphere is in hydrostatic equilibrium and the gradient of temperature follows the adiabatic approximation. However it has been proven by \citet{tatarski1971} that Eq.\ref{ann1:eq1} is more general and formally correct and can be obtained replacing T with the potential temperature $\theta$ that, differently from the temperature, is a conservative quantity. In some parts of the atmosphere (typically in the high part of the atmosphere) the difference can not be negligible. In reality the difference between Eq.\ref{ann1:eq1} and Eq.\ref{ann1:eq2} has negligible impact in our context because the model has to be calibrated and the calibration makes the impact on the model outputs irrelevant \citep{masciadri2015}. We prefer, however, to use Eq.\ref{ann1:eq1} because is more general and formally correct. This expression can be found, correctly used, in other studies (for example \citet{coulman1988,abahamid2004,cherubini2013}). Therefore it has not to be considered as a novelty but just a warning to avoid in the future a not suitable use of the Gladstone's relation.
%
%
\section{Correlation coefficient}
\label{annex_b}

\begin{figure}
\begin{center}
\begin{tabular}{c}
\includegraphics[width=0.9\columnwidth]{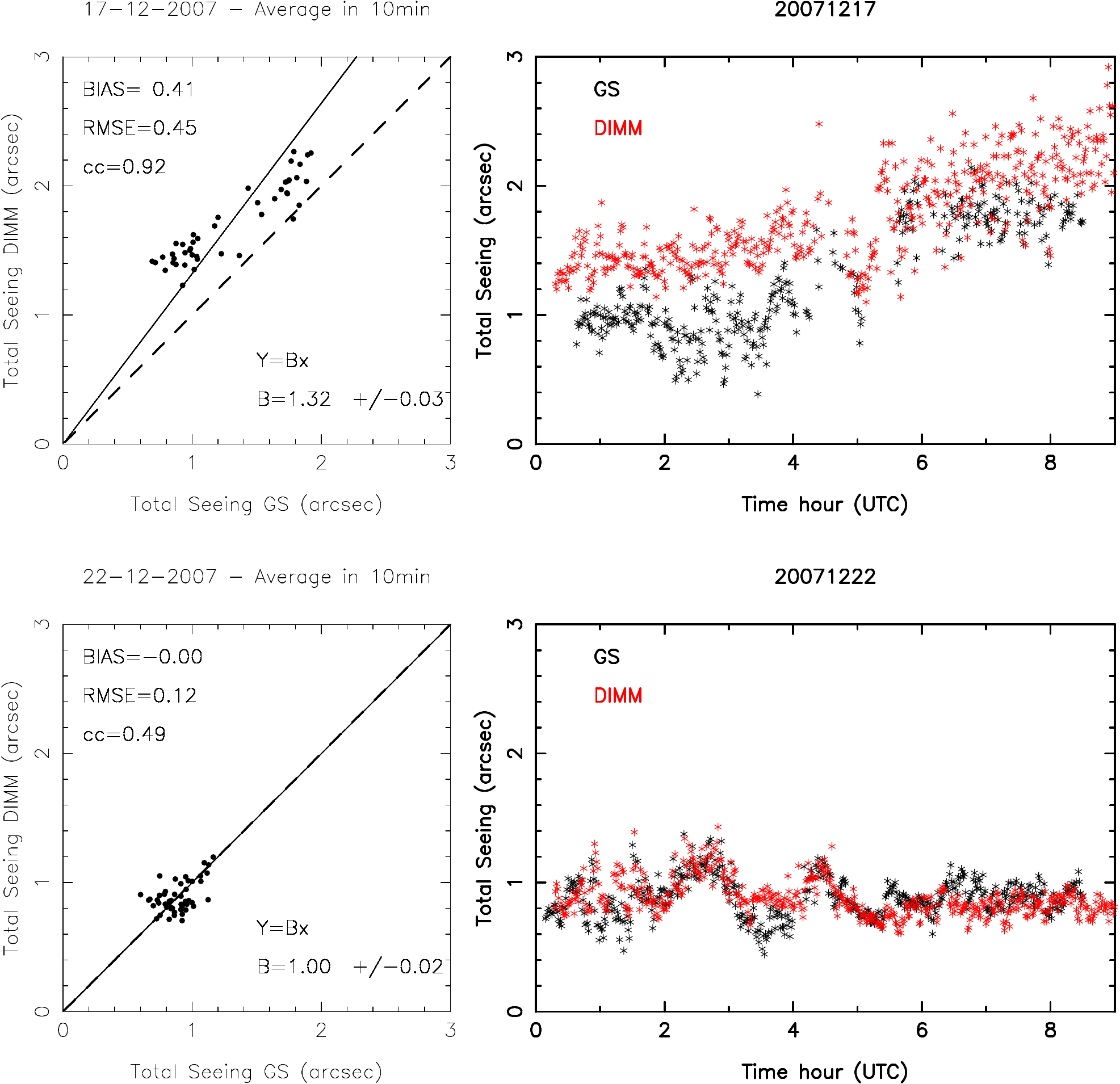}
\end{tabular}
\end{center}
\caption{\label{fig:example_single_cc} Example of the temporal evolutions of the total seeing (right) for two different
nights between GS and DIMM and the corresponding values of bias, RMSE and correlation coefficient (cc) (left).}
\end{figure}

The statistical estimators used for this analysis are: bias, RMSE and regression line passing by the origin. 
We decided not to include the correlation coefficient (cc) because, after a detailed analysis, we realised that the cc is not an estimator 
as reliable as others and, in our context, it can lead us to misleading conclusions with respect to our scientific goal. 
Just as an example Fig.~\ref{fig:example_single_cc} shows the temporal evolution of the total seeing obtained with GS and DIMM for two different nights. 
On the left side are reported the bias, RMSE and cc in the scattered plot. 
The cc of the night 17/12/2007 is 0.92 while the cc of the night 22/12/2007 is 0.49. 
However, looking at Fig.~\ref{fig:example_single_cc} (right side) it is evident that on 22/12/2007 the model has a better performance than on 17/12/2007 in spite of a cc=0.49 (instead of 0.92). What is important for us is how much close or far to observations are the values calculated by the model during the night. Moreover, we note that the cc does not tell us anything about the temporal trend of the estimate. 
We concluded therefore that the cc does not provide any further useful information with respect to bias, RMSE and regression line and, on the contrary, 
risks to induce to misleading conclusions. 

%
%
\section{Definitions of: Contingency Tables, PC, POD and EBD}
\label{annex_c}

Table \ref{tab:ct_gen33} is an example of a generic 3$\times$3 contingency table where the observations and simulations are divided into three categories delimited by two thresholds. PC, POD$_{i}$ and EBD can be defined using {\it a},{\it b},{\it c},{\it d},{\it e},{\it f},{\it g},{\it h},{\it i} (number of times in which an observation and a simulation fall inside each category) and N (the total events). The percentage of correct detection PC is defined in Eq.\ref{eq:pc2} where PC = 100\% is the best score; the probability to detect the value of a parameter inside a specific range of values (POD$_{i}$) is given by Eq.\ref{eq:pod1}-Eq.\ref{eq:pod3} where POD$_{i}$ = 100\% is the best score. The extremely bad detection (EBD) probability is given by Eq.\ref{eq:ebd} where EBD = 0\% is the best score. For a total random prediction and in case of a 3$\times$3 contingency table we have {\it a} = {\it b} = ... = {\it i} = {\it N/9} and PC = POD$_{i}$ = 33\% and EBD = 22.2\%.

\begin{equation}
PC=\frac{a+e+i}{N} \times 100; 0\%\leq PC\leq 100\%
\label{eq:pc2}
\end{equation}
\begin{equation}
POD(event_1)=\frac{a}{a+d+g} \times 100; 0\%\leq POD\leq 100\%
\label{eq:pod1}
\end{equation}
\begin{equation}
POD(event_2)=\frac{e}{b+e+h} \times 100; 0\%\leq POD\leq 100\%
\label{eq:pod2}
\end{equation}
\begin{equation}
POD(event_3)=\frac{i}{c+f+i} \times 100; 0\%\leq POD\leq 100\%
\label{eq:pod3}
\end{equation}
\begin{equation}
EBD=\frac{c+g}{N} \times 100; 0\%\leq EBD\leq 100\%
\label{eq:ebd}
\end{equation}

\begin{table}
\begin{center}
\caption{\label{tab:ct_gen33} Generic 3$\times$3 contingency table.}
\begin{adjustbox}{max width=\columnwidth}
\begin{tabular}{cccccc}
\multicolumn{2}{c}{\multirow{2}{*}{Intervals}} & \multicolumn{3}{c}{\bf OBSERVATIONS} &\\
\multicolumn{2}{c}{ } & 1 & 2 & 3 & Total \\
\hline
\multirow{10}{*}{\rotatebox{90}{\bf MODEL}} & & &  &\\
 & \multirow{2}{*}{1}  & a  &  \multirow{2}{*}{b}  &  \multirow{2}{*}{c} &  a+b+c \\
 &           & (hit 1) & & &  1 (Model)\\
 &      &    &    &  &  \\
 & \multirow{2}{*}{2}  & \multirow{2}{*}{d}  & e & \multirow{2}{*}{f} & d+e+f\\
 &           &    & (hit 2) & & 2 (Model)\\
 &      &    &   &  &   \\
 & \multirow{2}{*}{3}  & \multirow{2}{*}{g}  & \multirow{2}{*}{h} & i & g+h+i\\
 &           &    &  & (hit 3) & 3 (Model)\\
 &      &    &   &  &   \\
\hline
 & \multirow{2}{*}{Total} & a+d+g           &  b+e+h        & c+f+i & N=a+b+c+d+e+f+g+h+i \\
 &                        & 1 (OBS) & 2 (OBS) & 3 (OBS) & Total of events \\
\end{tabular}
\end{adjustbox}
\end{center}
\end{table}

\FloatBarrier
%
%
\section{Contingency Tables - Results}
\label{annex_d}

To facilitate the reading of the paper we gathered in this Annex all the contingency tables useful to summarise and discuss results presented in Section \ref{results}.

%
%
\begin{table}
\begin{center}
\caption{\label{tab:ct_seeing_val_2010_2} Sample (A): Contingency tables for the total seeing, in summer (42 nights in
2010 and 2011) - validation sample (CASE 1). 
Rounded values of the first and second tertiles of the climatology have been used as thresholds.}
\begin{adjustbox}{max width=\columnwidth}
\begin{tabular}{ccccc}
\multicolumn{2}{c}{ $\varepsilon$ - SUMMER} & \multicolumn{3}{c}{\bf DIMM}\\
\multicolumn{2}{c}{42 nights over 89} & $ \varepsilon< 0.88 $ & $ 0.88 < \varepsilon < 1.17 $   &  $ \varepsilon > 1.17$ \\
\multicolumn{2}{c}{(")} &   &    &   \\
\hline
\multirow{7}{*}{\rotatebox{90}{\bf MNH}} & & &\\
& $ \varepsilon < 0.88$                     & 181 & 175 & 125  \\
 &      &  &  & \\
 & $0.88 < \varepsilon <1.17 $ & 129 & 106 & 83 \\
 &      &  &  & \\
 & $ \varepsilon > 1.17$                   & 66 & 95 & 168  \\
 &      &  &  & \\
\hline
\multicolumn{5}{l}{Total points =  1128; $PC$=40.34\%; $EBD$=16.93\%} \\
\multicolumn{5}{l}{$POD_1$=48.14\%; $POD_2$=28.19\%; $POD_3$=44.68\%} \\
\end{tabular}
\end{adjustbox}
\end{center}
\end{table}
\begin{table}
\begin{center}
\caption{\label{tab:ct_seeing_val_2010_1} Sample (A): Contingency tables for the total seeing, in summer (42 nights in 2010 and 2011) - validation sample (CASE 2).}
\begin{adjustbox}{max width=\columnwidth}
\begin{tabular}{ccccc}
\multicolumn{2}{c}{ $\varepsilon$ - SUMMER} & \multicolumn{3}{c}{\bf DIMM }\\
\multicolumn{2}{c}{42 nights over 89} & $ \varepsilon < 1 $ & $ 1 < \varepsilon < 1.4$   &  $ \varepsilon > 1.4$ \\
\multicolumn{2}{c}{(")} &  &    &  \\
\hline
\multirow{7}{*}{\rotatebox{90}{\bf MNH }} & & &\\
& $ \varepsilon < 1$                     & 353 & 209 & 77  \\
 &      &  &  & \\
 & $1 < \varepsilon <1.4 $ & 177 & 77 & 43 \\
 &      &  &  & \\
 & $ \varepsilon> 1.4$                   & 39 & 48 & 105  \\
 &      &  &  & \\
\hline
\multicolumn{5}{l}{Total points =  1128; $PC$=47.43\%; $EBD$=10.28\%} \\
\multicolumn{5}{l}{$POD_1$=62.04\%; $POD_2$=23.05\%; $POD_3$=46.67\%} \\
\end{tabular}
\end{adjustbox}
\end{center}
\end{table}
\begin{table}
\begin{center}
\caption{\label{tab:ct_seeing_val_2007_2} Sample (B): Contingency tables for the total seeing, in summer (18 nights in 2007) - validation sample (CASE 1). 
Rounded values of the first and second tertiles of the climatology have been used as thresholds.}
\begin{adjustbox}{max width=\columnwidth}
\begin{tabular}{ccccc}
\multicolumn{2}{c}{ $\varepsilon$ - SUMMER} & \multicolumn{3}{c}{\bf DIMM}\\
\multicolumn{2}{c}{18 nights over 36} & $ \varepsilon< 0.86 $ & $ 0.86 < \varepsilon < 1.20 $   &  $ \varepsilon > 1.20$ \\
\multicolumn{2}{c}{(")} & &   &   \\
\hline
\multirow{7}{*}{\rotatebox{90}{\bf MNH}} & & &\\
& $ \varepsilon < 0.86$                     & 86 & 87 & 52  \\
 &      &  &  & \\
 & $0.86 < \varepsilon <1.20 $ & 64 & 61 & 34 \\
 &      &  &  & \\
 & $ \varepsilon > 1.20$                   & 12 & 14 & 76  \\
 &      &  &  & \\
\hline
\multicolumn{5}{l}{Total points =  486; $PC$=45.88\%; $EBD$=13.17\%} \\
\multicolumn{5}{l}{$POD_1$=53.09\%; $POD_2$=37.65\%; $POD_3$=46.91\%} \\
\end{tabular}
\end{adjustbox}
\end{center}
\end{table}
\begin{table}
\begin{center}
\caption{\label{tab:ct_seeing_val_2007_1} Sample (B): Contingency tables for the total seeing, in summer (18 nights in 2007) - validation sample (CASE 2).}
\begin{adjustbox}{max width=\columnwidth}
\begin{tabular}{ccccc}
\multicolumn{2}{c}{ $\varepsilon$ - SUMMER} & \multicolumn{3}{c}{\bf DIMM}\\
\multicolumn{2}{c}{18 nights over 36} & $ \varepsilon< 1 $ & $ 1 < \varepsilon < 1.4$   &  $ \varepsilon > 1.4$ \\
\multicolumn{2}{c}{(")} &  &   &   \\
\hline
\multirow{7}{*}{\rotatebox{90}{\bf MNH}} & & &\\
& $ \varepsilon < 1$                     & 177 & 87 & 38  \\
 &      &  &  & \\
 & $1 < \varepsilon <1.4$ & 58 & 34 & 14 \\
 &      &  &  & \\
 & $ \varepsilon > 1.4$                   & 9 & 18 & 51  \\
 &      &  &  & \\
\hline
\multicolumn{5}{l}{Total points =  486; $PC$=53.91\%; $EBD$=9.67\%} \\
\multicolumn{5}{l}{$POD_1$=72.54\%; $POD_2$=24.46\%; $POD_3$=49.51\%} \\
\end{tabular}
\end{adjustbox}
\end{center}
\end{table}
%
%
%
\begin{table}
\begin{center}
\caption{\label{tab:ct_seeing_cal_gs_dimm_1} Contingency tables for the total seeing, between DIMM and GS, using GS as a reference - calibration sample (20 nights).}
\begin{adjustbox}{max width=\columnwidth}
\begin{tabular}{ccccc}
\multicolumn{2}{c}{$\varepsilon$} & \multicolumn{3}{c}{\bf GS }\\
\multicolumn{2}{c}{20 nights} & $ \varepsilon < 0.97 $ & $ 0.97 < \varepsilon < 1.24 $   &  $ \varepsilon > 1.24$ \\
\multicolumn{2}{c}{(")} &  &   &  \\
\hline
\multirow{7}{*}{\rotatebox{90}{\bf DIMM}} & & &\\
& $ \varepsilon < 0.97$                     & 108 & 45 & 5 \\
 &      &  &  & \\
 & $0.97 < \varepsilon <1.24 $ & 39 & 60 & 14 \\
 &      &  &  & \\
 & $ \varepsilon> 1.24$                   & 21 & 63 & 149  \\
 &      &  &  & \\
\hline
\multicolumn{5}{l}{Total points =  504; $PC$=62.90\%; $EBD$=5.16\%} \\
\multicolumn{5}{l}{$POD_1$=64.29\%; $POD_2$=35.71\%; $POD_3$=88.69\%} \\
\end{tabular}
\end{adjustbox}
\end{center}
\end{table}
\begin{table}
\begin{center}
\caption{\label{tab:ct_seeing_cal_gs_dimm_2} Contingency tables for the total seeing, between DIMM and GS, using DIMM as a reference - calibration sample (20 nights).}
\begin{adjustbox}{max width=\columnwidth}
\begin{tabular}{ccccc}
\multicolumn{2}{c}{$\varepsilon$} & \multicolumn{3}{c}{\bf DIMM }\\
\multicolumn{2}{c}{20 nights} & $ \varepsilon < 1 $ & $ 1 < \varepsilon < 1.42 $   &  $ \varepsilon > 1.42$ \\
\multicolumn{2}{c}{(")} & &    &   \\
\hline
\multirow{7}{*}{\rotatebox{90}{\bf GS}} & & &\\
& $ \varepsilon < 1$                     & 120 & 52 & 13 \\
 &      &  &  & \\
 & $1 < \varepsilon <1.42 $ & 48 & 103 & 77 \\
 &      &  &  & \\
 & $ \varepsilon> 1.42$                   & 0 & 13 & 78  \\
 &      &  &  & \\
\hline
\multicolumn{5}{l}{Total points =  504; $PC$=59.72\%; $EBD$=2.58\%} \\
\multicolumn{5}{l}{$POD_1$=71.43\%; $POD_2$=61.31\%; $POD_3$=46.43\%} \\
\end{tabular}
\end{adjustbox}
\end{center}
\end{table}
\begin{table}
\begin{center}
\caption{\label{tab:ct_seeing_cal_gs_mnh} Contingency tables for the total seeing, between Meso-NH and GS - calibration sample (20 nights).}
\begin{adjustbox}{max width=\columnwidth}
\begin{tabular}{ccccc}
\multicolumn{2}{c}{$\varepsilon$} & \multicolumn{3}{c}{\bf GS}\\
\multicolumn{2}{c}{20 nights} & $ \varepsilon< 0.97 $ & $ 0.97 < \varepsilon < 1.24 $   &  $ \varepsilon > 1.24$ \\
\multicolumn{2}{c}{(")} &  &    &  \\
\hline
\multirow{7}{*}{\rotatebox{90}{\bf MODEL}} & & &\\
& $ \varepsilon < 0.97$                     & 138 & 115 &15  \\
 &      &  &  & \\
 & $0.97 < \varepsilon <1.24 $ & 18  & 35 & 48 \\
 &      &  &  & \\
 & $ \varepsilon > 1.24$                   & 12 & 18 & 105  \\
 &      &  &  & \\
\hline
\multicolumn{5}{l}{Total points =  504; $PC$=55.16\%; $EBD$=5.36\%} \\
\multicolumn{5}{l}{$POD_1$=82.14\%; $POD_2$=20.84\%; $POD_3$=62.50\%} \\
\end{tabular}
\end{adjustbox}
\end{center}
\end{table}
\begin{table}
\begin{center}
\caption{\label{tab:ct_seeing_cal_dimm_mnh} Contingency tables for the total seeing, between Meso-NH and DIMM - calibration sample (20 nights).}
\begin{adjustbox}{max width=\columnwidth}
\begin{tabular}{ccccc}
\multicolumn{2}{c}{$\varepsilon$} & \multicolumn{3}{c}{\bf DIMM}\\
\multicolumn{2}{c}{20 nights} & $ \varepsilon< 1 $ & $ 1 < \varepsilon < 1.42 $   &  $ \varepsilon > 1.42$ \\
\multicolumn{2}{c}{(")} & &    &  \\
\hline
\multirow{7}{*}{\rotatebox{90}{\bf MODEL}} & & &\\
 & $ \varepsilon < 1$                     &  156&  102 & 27 \\
 &      &  &  & \\
 & $1 < \varepsilon <1.42$ & 17 & 60 & 70 \\
 &      &  &  & \\
 & $ \varepsilon > 1.42$                   & 4 & 14 & 79  \\
 &      &  &  & \\
\hline
\multicolumn{5}{l}{Total points =  529; $PC$=55.76\%; $EBD$=5.86\%} \\
\multicolumn{5}{l}{$POD_1$=88.13\%; $POD_2$=34.10\%; $POD_3$=44.89\%} \\
\end{tabular}
\end{adjustbox}
\end{center}
\end{table}
%
%
\begin{table}
\begin{center}
\caption{\label{tab:ct_mass_dimm_tauO_val} Contingency table of \tauO\ between MASS and DIMM, using the DIMM as a reference (48 nights in 2010/2011) - validation sample.}
\begin{adjustbox}{max width=\columnwidth}
\begin{tabular}{ccccc}
\multicolumn{2}{c}{$\tau_{0}  $} & \multicolumn{3}{c}{\bf DIMM}\\
\multicolumn{2}{c}{48 nights} & $ \tau_{0} < 2.07 $ & $ 2.07 < \tau_{0}  < 3.26 $   &  $ \tau_{0}  > 3.26$ \\
\multicolumn{2}{c}{(ms)} &  &    &   \\
\hline
\multirow{7}{*}{\rotatebox{90}{\bf MASS}} & & &\\
& $ \tau_{0}  < 2.07$                     & 182 & 15 & 8  \\
 &      &  &  & \\
 & $2.07 < \tau_{0}  <3.26 $ & 167 & 139 & 16 \\
 &      &  &  & \\
 & $ \tau_{0}  > 3.26$                   & 46 & 240 & 370  \\
 &      &  &  & \\
\hline
\multicolumn{5}{l}{Total points =  1183; $PC$=58.41\%; $EBD$=4.56\%} \\
\multicolumn{5}{l}{$POD_1$=46.08\%; $POD_2$=35.28\%; $POD_3$=93.91\%} \\
\end{tabular}
\end{adjustbox}
\end{center}
\end{table}

\begin{table}
\begin{center}
\caption{\label{tab:ct_dimm_mass_tauO_val} Contingency table of \tauO\ between MASS and DIMM, using the MASS as a reference (48 nights in 2010/2011) - validation sample.}
\begin{adjustbox}{max width=\columnwidth}
\begin{tabular}{ccccc}
\multicolumn{2}{c}{$\tau_{0}  $} & \multicolumn{3}{c}{\bf MASS}\\
\multicolumn{2}{c}{48 nights} & $ \tau_{0} < 2.68 $ & $ 2.68 < \tau_{0}  < 4.87 $   &  $ \tau_{0}  > 4.87$ \\
\multicolumn{2}{c}{(ms)} &  &   &   \\
\hline
\multirow{7}{*}{\rotatebox{90}{\bf DIMM}} & & &\\
& $ \tau_{0}  < 2.68$                     & 364 & 215 &  65 \\
 &      &  &  & \\
 & $2.68 < \tau_{0}  < 4.87 $ & 20 & 170 & 226 \\
 &      &  &  & \\
 & $ \tau_{0}  > 4.87$                   & 11 & 9 & 103  \\
 &      &  &  & \\
\hline
\multicolumn{5}{l}{Total points =  1183; $PC$=53.85\%; $EBD$=6.42\%} \\
\multicolumn{5}{l}{$POD_1$=92.15\%; $POD_2$=43.15\%; $POD_3$=26.14\%} \\
\end{tabular}
\end{adjustbox}
\end{center}
\end{table}
\begin{table}
\begin{center}
\caption{\label{tab:ct_dimm_mnh_summer_tauO_val} Contingency table of \tauO\ between Meso-NH and DIMM, in summer - validation sample.}
\begin{adjustbox}{max width=\columnwidth}
\begin{tabular}{ccccc}
\multicolumn{2}{c}{$\tau_{0} - SUMMER $} & \multicolumn{3}{c}{\bf DIMM}\\
\multicolumn{2}{c}{42 nights over 89} & $ \tau_{0} < 2.20 $ & $ 2.20 < \tau_{0}  < 3.58 $   &  $ \tau_{0}  > 3.58$ \\
\multicolumn{2}{c}{(ms)} &  &    &  \\
\hline
\multirow{7}{*}{\rotatebox{90}{\bf MNH}} & & &\\
& $ \tau_{0}  < 2.20$                     & 178 & 61 & 13  \\
 &      &  &  & \\
 & $2.20 < \tau_{0}  <3.58 $ & 138 & 139 & 95 \\
 &      &  &  & \\
 & $ \tau_{0}  > 3.58$                   & 54 & 169 & 261  \\
 &      &  &  & \\
\hline
\multicolumn{5}{l}{Total points =  1108; $PC$=57.17\%; $EBD$=6.05\%} \\
\multicolumn{5}{l}{$POD_1$=48.11\%; $POD_2$=37.67\%; $POD_3$=70.73\%} \\
\end{tabular}
\end{adjustbox}
\end{center}
\end{table}
\begin{table}
\begin{center}
\caption{\label{tab:ct_dimm_mnh_winter_tauO_val} Contingency table of \tauO\ between Meso-NH and DIMM, in winter - validation sample.}
\begin{adjustbox}{max width=\columnwidth}
\begin{tabular}{ccccc}
\multicolumn{2}{c}{$\tau_{0} - WINTER $} & \multicolumn{3}{c}{\bf DIMM}\\
\multicolumn{2}{c}{47 nights over 89} & $ \tau_{0} < 1.67 $ & $ 1.67 < \tau_{0}  < 2.48 $   &  $ \tau_{0}  > 2.48$ \\
\multicolumn{2}{c}{(ms)} &  &   &   \\
\hline
\multirow{7}{*}{\rotatebox{90}{\bf MNH}} & & &\\
& $ \tau_{0}  < 1.67$                     & 271 & 187 & 37  \\
 &      &  &  & \\
 & $1.67 < \tau_{0}  <2.48 $ & 118 & 116 & 107 \\
 &      &  &  & \\
 & $ \tau_{0}  > 2.48$                   & 31 & 116 & 275  \\
 &      &  &  & \\
\hline
\multicolumn{5}{l}{Total points =  1258; $PC$=52.62\%; $EBD$=5.40\%} \\
\multicolumn{5}{l}{$POD_1$=64.52\%; $POD_2$=27.68\%; $POD_3$=65.63\%} \\
\end{tabular}
\end{adjustbox}
\end{center}
\end{table}
\begin{table}
\begin{center}
\caption{\label{tab:ct_mass_mnh_summer_tauO_val} Contingency table of \tauO\ between Meso-NH and MASS, in summer - validation sample.}
\begin{adjustbox}{max width=\columnwidth}
\begin{tabular}{ccccc}
\multicolumn{2}{c}{$\tau_{0} - SUMMER $} & \multicolumn{3}{c}{\bf MASS}\\
\multicolumn{2}{c}{21 nights over 48} & $ \tau_{0} < 3.27 $ & $ 3.27 < \tau_{0}  < 5.36 $   &  $\tau_{0}  > 5.36$ \\
\multicolumn{2}{c}{(ms)} &  &    &   \\
\hline
\multirow{7}{*}{\rotatebox{90}{\bf MNH}} & & &\\
& $ \tau_{0}  < 3.27$                     & 111 & 70 & 7  \\
 &      &  &  & \\
 & $3.27 < \tau_{0}  < 5.36 $ & 49& 52 &38 \\
 &      &  &  & \\
 & $ \tau_{0}  > 5.36$                   & 1 & 38 & 115  \\
 &      &  &  & \\
\hline
\multicolumn{5}{l}{Total points =  481; $PC$=57.80\%; $EBD$=1.66\%} \\
\multicolumn{5}{l}{$POD_1$=68.94\%; $POD_2$=32.50\%; $POD_3$=71.87\%} \\
\end{tabular}
\end{adjustbox}
\end{center}
\end{table}
\begin{table}
\begin{center}
\caption{\label{tab:ct_mass_mnh_winter_tauO_val} Contingency table of \tauO\ between Meso-NH and MASS, in winter - validation sample.}
\begin{adjustbox}{max width=\columnwidth}
\begin{tabular}{ccccc}
\multicolumn{2}{c}{$\tau_{0} - WINTER $} & \multicolumn{3}{c}{\bf MASS}\\
\multicolumn{2}{c}{27 nights over 48} & $ \tau_{0} < 2.45 $ & $ 2.45 < \tau_{0}  < 4.19 $   &  $\tau_{0}  > 4.19$ \\
\multicolumn{2}{c}{(ms)} &  &   &   \\
\hline
\multirow{7}{*}{\rotatebox{90}{\bf MNH}} & & &\\
& $ \tau_{0}  < 2.45$                     & 112 & 86 & 6  \\
 &      &  &  & \\
 & $2.45 < \tau_{0}  < 4.19 $ & 94& 108 & 45 \\
 &      &  &  & \\
 & $ \tau_{0}  > 4.19$                   & 28 & 40 & 183  \\
 &      &  &  & \\
\hline
\multicolumn{5}{l}{Total points =  702; $PC$=57.41\%; $EBD$=4.84\%} \\
\multicolumn{5}{l}{$POD_1$=47.86\%; $POD_2$=46.15\%; $POD_3$=78.20\%} \\
\end{tabular}
\end{adjustbox}
\end{center}
\end{table}
%
%
\begin{table}
\begin{center}
\caption{\label{tab:ct_dimm_mass_thetaO_2010} Contingency table of \thetaO\ between MASS and DIMM, using the MASS as a reference (44 nights in 2010/2011).}
\begin{adjustbox}{max width=\columnwidth}
\begin{tabular}{ccccc}
\multicolumn{2}{c}{$\theta_{0}  $} & \multicolumn{3}{c}{\bf MASS}\\
\multicolumn{2}{c}{44 nights} & $ \theta_{0}< 1.53$ & $ 1.53 < \theta_{0} < 2.04 $   &  $ \theta_{0} > 2.04$ \\
\multicolumn{2}{c}{(")} &  &    &  \\
\hline
\multirow{7}{*}{\rotatebox{90}{\bf DIMM}} & & &\\
& $ \theta_{0} < 1.53$                     & 47 & 1 &  0 \\
 &      &  &  & \\
 & $1.53 < \theta_{0} < 2.04 $ & 157 & 45 & 17 \\
 &      &  &  & \\
 & $ \theta_{0} > 2.04$                   & 168 & 325 & 354  \\
 &      &  &  & \\
\hline
\multicolumn{5}{l}{Total points =  1114; $PC$=40.04\%; $EBD$=15.08\%} \\
\multicolumn{5}{l}{$POD_1$=12.63\%; $POD_2$=12.13\%; $POD_3$=95.42\%} \\
\end{tabular}
\end{adjustbox}
\end{center}
\end{table}
\begin{table}
\begin{center}
\caption{\label{tab:ct_mass_dimm_thetaO_2010} Contingency table of \thetaO\ between MASS and DIMM, using the DIMM as a reference (44 nights in 2010/2011).}
\begin{adjustbox}{max width=\columnwidth}
\begin{tabular}{ccccc}
\multicolumn{2}{c}{$\theta_{0}  $} & \multicolumn{3}{c}{\bf DIMM}\\
\multicolumn{2}{c}{44 nights} &  $ \theta_{0}   < 2.26 $ & $  2.26 < \theta_{0}  < 2.94 $   &  $ \theta_{0}  > 2.94$ \\
\multicolumn{2}{c}{(")} &   &   &   \\
\hline
\multirow{7}{*}{\rotatebox{90}{\bf MASS}} & & &\\
& $ \theta_{0}  < 2.26$                     & 354 & 329 &  177 \\
 &      &  &  & \\
 & $2.26 < \theta_{0}  < 2.94 $ & 13 & 38 & 156 \\
 &      &  &  & \\
 & $ \theta_{0}  > 2.94$                   & 5 & 4 & 38  \\
 &      &  &  & \\
\hline
\multicolumn{5}{l}{Total points =  1114; $PC$=38.60\%; $EBD$=16.34\%} \\
\multicolumn{5}{l}{$POD_1$=95.16\%; $POD_2$=10.24\%; $POD_3$=10.24\%} \\
\end{tabular}
\end{adjustbox}
\end{center}
\end{table}
\begin{table}
\begin{center}
\caption{\label{tab:ct_mass_mnh_thetaO_2010_summer} Contingency table of \thetaO\ between MASS and Meso-NH (21 summer nights in 2010/2011)}
\begin{adjustbox}{max width=\columnwidth}
\begin{tabular}{ccccc}
\multicolumn{2}{c}{$\theta_{0} - SUMMER $} & \multicolumn{3}{c}{\bf MASS}\\
\multicolumn{2}{c}{21 nights over 48} &  $ \theta_{0}   < 1.55 $ & $  1.55 < \theta_{0}  < 2.01 $   &  $ \theta_{0}  > 2.01$ \\
\multicolumn{2}{c}{(")} &   &   &   \\
\hline
\multirow{7}{*}{\rotatebox{90}{\bf MNH}} & & &\\
& $ \theta_{0}  < 1.55$                     & 16& 18 &  12 \\
 &      &  &  & \\
 & $1.55 < \theta_{0}  < 2.05 $ & 88 & 80 & 61 \\
 &      &  &  & \\
 & $ \theta_{0}  > 2.05$                   & 70  & 75 & 100  \\
 &      &  &  & \\
\hline
\multicolumn{5}{l}{Total points =  520; $PC$=37.69\%; $EBD$=15.77\%} \\
\multicolumn{5}{l}{$POD_1$=9.20\%; $POD_2$=46.24\%; $POD_3$=57.80\%} \\
\end{tabular}
\end{adjustbox}
\end{center}
\end{table}

\clearpage
%
%
\section{Optical turbulence temporal evolution}
\label{annex_e}

\begin{figure*}
\begin{center}
\begin{tabular}{c}
\includegraphics[width=0.55\textwidth]{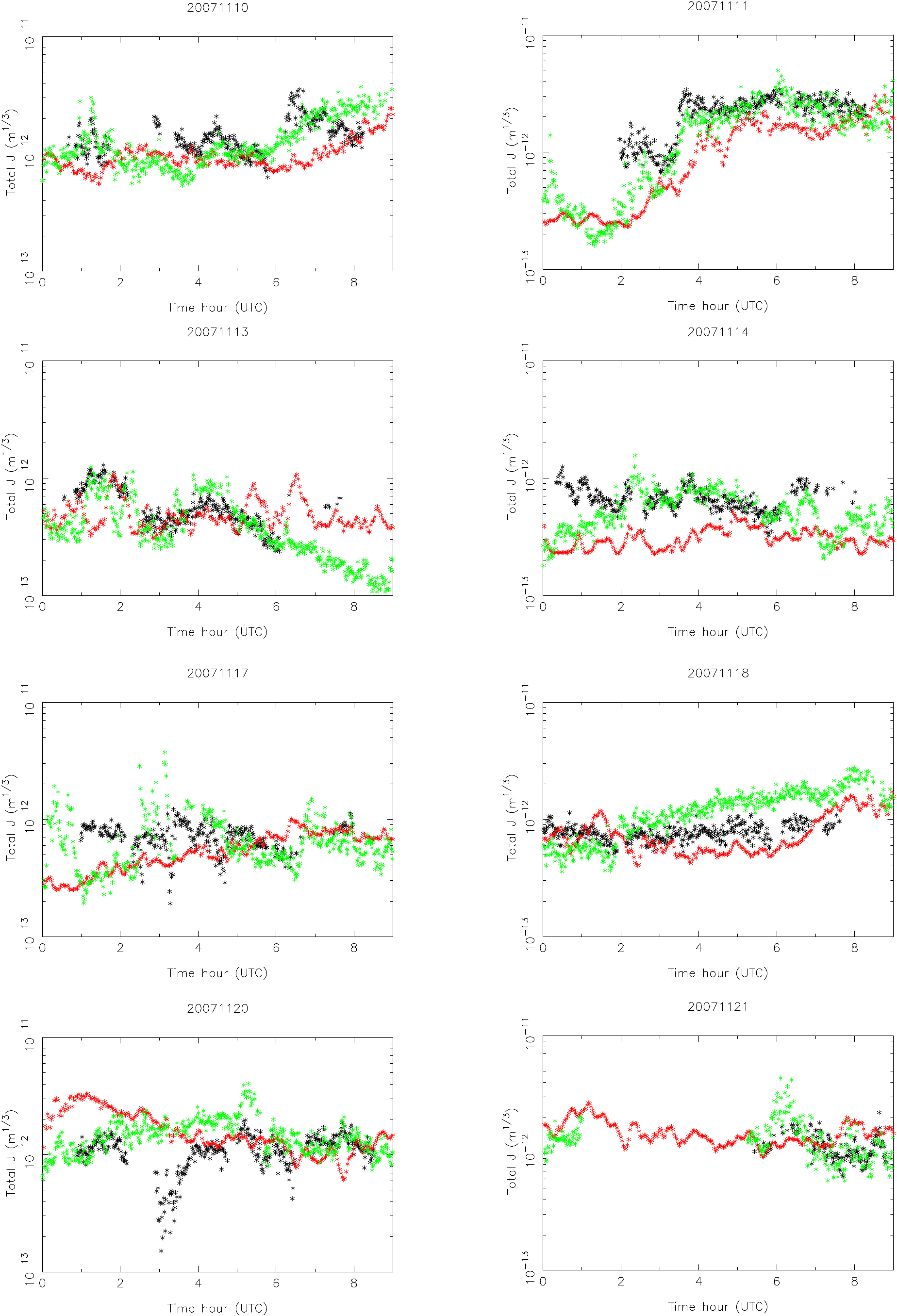}
\end{tabular}
\end{center}
\caption{\label{fig:temp_evol_J1} Temporal evolutions of J for all the 20 nights of the PAR2007 site testing campaign: GS (black), DIMM (green) and Astro-Meso-Nh model (red).}
\end{figure*}

\begin{figure*}
\begin{center}
\begin{tabular}{c}
\includegraphics[width=0.55\textwidth]{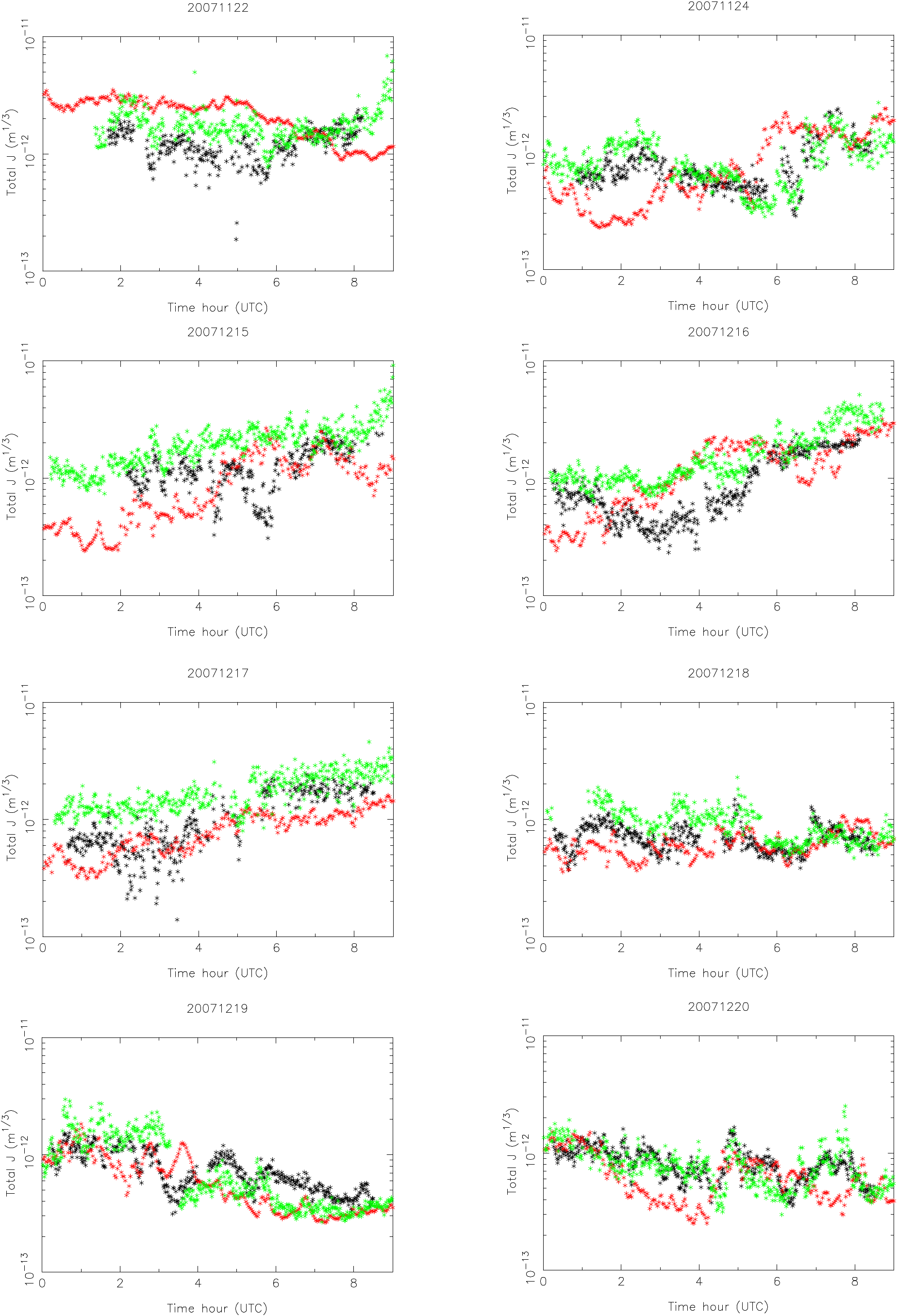}
\end{tabular}
\end{center}
\caption{\label{fig:temp_evol_J2} As Fig.\ref{fig:temp_evol_J1} - continued.}
\end{figure*}

\begin{figure*}
\begin{center}
\begin{tabular}{c}
\includegraphics[width=0.55\textwidth]{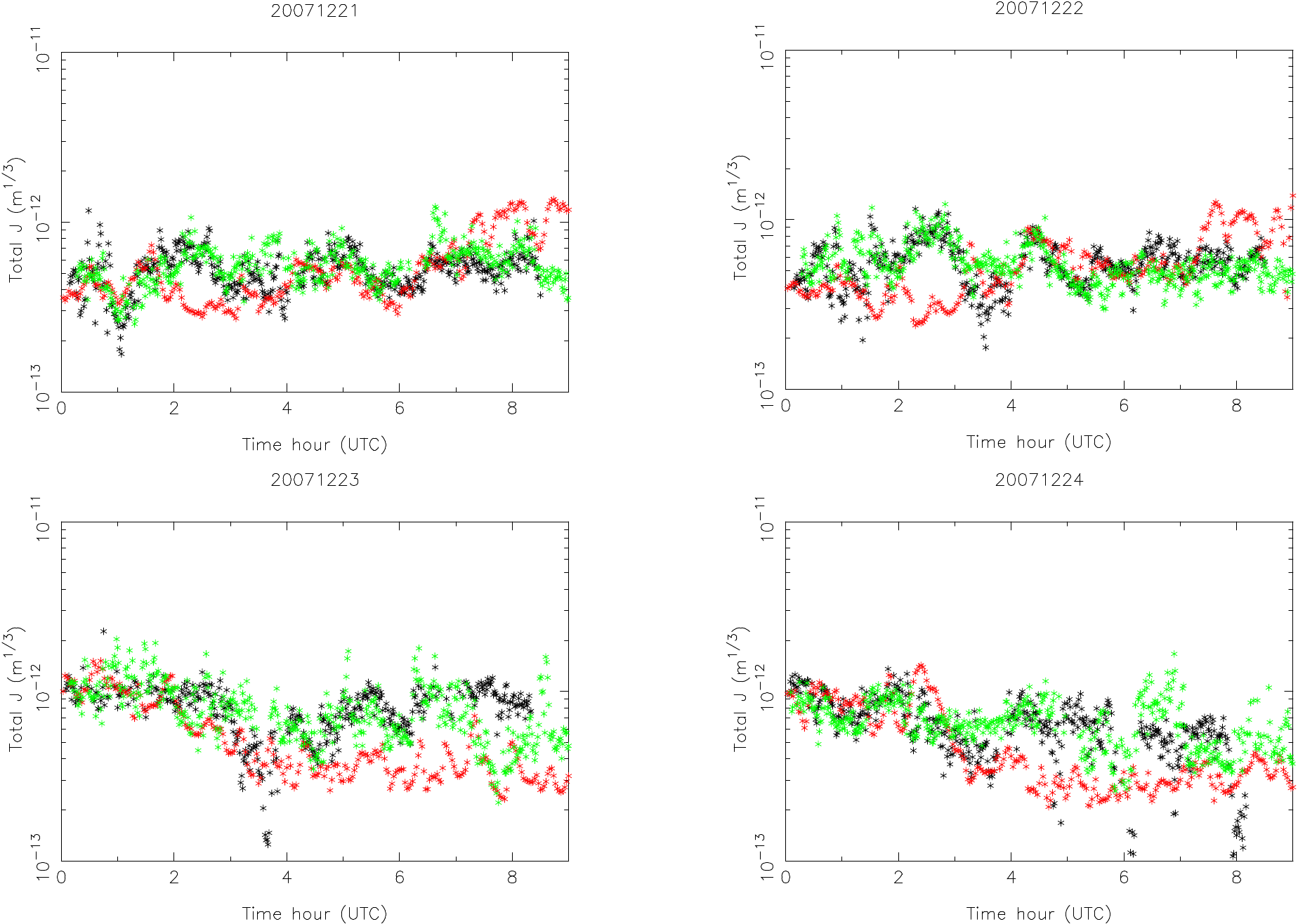}
\end{tabular}
\end{center}
\caption{\label{fig:temp_evol_J3} As Fig.\ref{fig:temp_evol_J1} - continued.}
\end{figure*}

Fig.\ref{fig:temp_evol_J1}-Fig.\ref{fig:temp_evol_J3} show the temporal evolution of the optical turbulence developed on the whole atmosphere ($\sim$ 20 km) during all 20 nights of the PAR2007 site testing campaign. Astro-Meso-Nh model estimates are compared to observations done with a GS and a DIMM.
%
%
%

%



\bsp	
\label{lastpage}
\end{document}